\documentclass[twocolumn,showpacs,preprintnumbers,amsmath,amssymb,prb,aps]{revtex4-1}
\usepackage{epsfig}
\usepackage{epstopdf}
\usepackage{multirow}
\usepackage{graphicx}
\usepackage{dcolumn}
\usepackage{bm}
\usepackage{textcomp}
\usepackage{array}
\usepackage{caption}
\usepackage{xfrac}
\usepackage{mathtools}
\usepackage{amsmath}
\usepackage{subfigure}
\usepackage{hyperref}
\hypersetup{
    colorlinks=true,
    linkcolor=blue,
    filecolor=magenta,      
    urlcolor=blue,
    pdftitle={},
    }

\begin{document}

\title{Existence of \textit{nodal-arc} and its evolution into Weyl-nodes in the presence of spin-orbit coupling in TaAs \& TaP.}

\author{Vivek Pandey$^{1}$ }
\altaffiliation{vivek6422763@gmail.com}
\author{Sudhir K. Pandey$^{2}$}
\altaffiliation{sudhir@iitmandi.ac.in}
\affiliation{$^{1}$School of Physical Sciences, Indian Institute of Technology Mandi, Kamand - 175075, India\\
$^{2}$School of Mechanical and Materials Engineering, Indian Institute of Technology Mandi, Kamand - 175075, India}

\date{\today}

\begin{abstract}

 In this work, we report the existence of \textit{nodal-arc}, which acts as the building block of all the nodal-rings in TaAs \& TaP. This \textit{nodal-arc} is found to be capable of generating all the nodal-rings in these materials upon the application of space-group symmetry operations including time-reversal symmetry. The arcs are obtained to be dispersive with the energy spread of $\sim$109 ($\sim$204) meV in TaAs (TaP). Also, the orbitals leading to bands-inversion and thus the formation of \textit{nodal-arcs} are found to be Ta-5\textit{d} \& As-4\textit{p} (P-3\textit{p}) in TaAs (TaP). The area of nodal-rings is found to be highly sensitive to the change in hybridization-strength, where the increase in hybridization-strength leads to the decrease in the area of nodal-rings. In the presence of spin-orbit coupling (SOC), all the points on these arcs get gaped-up and two pairs of Weyl-nodes are found to evolve from them. Out of the two pair, one is found to be situated close to the joining point of the two arcs forming a ring. This causes the evolution of each nodal-ring into three pairs of Weyl-nodes. The coordinates of these Weyl-nodes are found to be robust to the increase in SOC-strength from $\sim$ 0.7-3.5 eV. All the results are obtained at the \textit{first-principle} level. This work provides a clear picture of the existence of \textit{nodal-arc} due to accidental degeneracy and its evolution into Weyl-nodes under the effect of SOC.

\end{abstract}

\maketitle

\section{Introduction} 
\setlength{\parindent}{3em}

Topological bands-crossings near the Fermi level are of great importance in present condensed-matter physics research. The popular classes of materials associated with such bands-degeneracy includes- Weyl semimetals (WSMs)\cite{weng2015weyl,soluyanov2015type,lv2015experimental,pandey2021anab}, Dirac-semimetals (DSMs)\cite{neupane2014observation,steinberg2014bulk,yang2014classification} and nodal-line semimetals (NLSMs)\cite{burkov2011topological,yu2015topological,xu2017topological}. The WSMs contain discrete set of two-fold bands-intersections between conduction and valence bands across their full Brillouin zone (BZ). They are associated with special Fermi-arc surface states\cite{belopolski2016criteria,hosur2012friedel}. The DSMs possess discrete set of four-fold conduction and valence bands-touchings in their full BZ. The NLSMs are characterized by the bands-crossings near the Fermi-level leading to the formation of one-dimensional loop of bands-degeneracy. Such loops are popularly referred as nodal-lines, nodal-loops or nodal-rings. They are characterized by special drumbhead-like surface states enclosed within the nodal-lines. Apart from electronic states, topological bands-touchings are also reported in phononic states of several materials\cite{sihi2022evidence,miao2018observation}. Generally, bands-touchings in WSMs and NLSMs are due to accidental degeneracy which arise from the profile of the potential associated with these materials\cite{RevModPhys.90.015001,quan2017single}. 

Another important aspect regarding these materials is the number of such nodal-lines or Weyl-nodes which are present in their full Brillouin zone (BZ). In this regard, it also becomes important to find-out the minimal accidental degeneracy which can generate all the bands-touching points in the full BZ upon applying the crystal symmetry operations associated with the given material. This minimal accidental degeneracy can be associated with the irreducible parts of the BZ of the given semimetals. This part of degeneracy must be considered from fundamental degeneracy. The rest of the degenerate points in the full BZ are the consequences of the symmetry operations in the reciprocal space and hence, can be considered as derived one. In literature, a nodal-loop is typically presented as a fundamental entity. If a material contains more than one loop, the one of them is generally considered as a fundamental entity and the rest of the loops are found to be related through symmetry operations. However, one cannot rule out the possibility that within one loop, a section of it is an independent entity arising from fundamental degeneracy and the rest part of it can be derived from crystal symmetry. To the best of our knowledge, such discussion regarding a nodal-loop is not yet been reported. It is important to mention here that Dirac node arcs are reported to be observed in certain materials\cite{wu2016dirac}. In such works, Dirac node arc stands for Dirac-like bands degeneracy along a line which is not a closed loop. Such arcs are reported to be entirely a different entity and is noway related with nodal-loop. Thus, \textit{nodal-arc} term as used in this work is entirely different from Dirac node arcs.

Now, we shall discuss some of the mechanisms leading to the formation of nodal-lines. The well-known conditions leading to their formation in the absence of SOC includes- (i) due to band inversion of opposite parity bands\cite{kim2015dirac}, (ii) because of the simultaneous presence of inversion symmetry and time-reversal symmetry (TR)\cite{yu2015topological} and (iii) due to the presence of mirror symmetry and TR symmetry\cite{yang2014dirac,chiu2014classification}. Furthermore, in the presence of SOC, the combination of mirror symmetry and TR symmetry may also result in the formation of nodal-lines\cite{bian2016topological}. In several cases, under the effect of SOC, nodal-lines are reported to evolve into different topological states, such as- WSM\cite{huang2015weyl,weng2015weyl}, DSM\cite{tang2016dirac} or NLSM\cite{chen2017dirac,chen2015topological,bzduvsek2016nodal} with a new set of nodal-lines. Apart from the SOC-strength, new sets of degeneracy also depends upon the crystal symmetries and orbital character. Also, the nodal-loop may either form at constant energy or its different part may be situated at different energy (dispersive in nature)\cite{yang2018quantum}.  

Now we explore the class of materials which show nodal-lines forming due to band-inversion and under the effect of SOC, these nodal-lines are evolving into Weyl-nodes. Some of the candidate materials of this class include- (i) Co$_3$Sn$_2$S$_2$, in which there are several band-inversions centered at the \textit{L}-points resulting into the formation of several nodal-rings located at the mirror planes\cite{liu2018giant}. With the action of $C_{3z}$-rotation and inversion symmetries of the material, a total of six nodal rings is reported in its BZ. Each of these nodal-rings are extended beyond the boundaries of first BZ. Moving further, it is shown that when SOC is taken into account, each nodal rings are gaped out at all points except two, which form a pair of Weyl-nodes. These Weyl-nodes are situated at $\sim$60 meV above Fermi-energy. Also, out of the two Weyl-nodes generated from a nodal-ring, one is within the first BZ while the other is outside it. (ii) HfC, which is reported to contain two types of nodal-rings encircling the $\Gamma$ and $M$ points, respectively\cite{yu2017nodal}. One ring is encircling the $\Gamma$-point and 6 rings, which are related with the space-group symmetry of the material, are encircling the six $M$-points of the BZ. Thus, the ring encircling the $\Gamma$-point and the ring enclosing one of the $M$ point can be considered as related to fundamental degeneracy, as discussed in present work. The ring encircling the $\Gamma$-point is perpendicular to the plane containing the rings encircling the $M$-points. Also, ring encircling the $\Gamma$-point passes through each of the rings enclosing the $M$-points to form a structure popularly known as nodal-chain. The work also shows that the nodal-ring encircling the $M$-point is extended beyond the boundaries of the first BZ. When SOC is taken into account, the ring encircling the $\Gamma$-point evolve into 6 pairs of Weyl-nodes while the ring enclosing one of the $M$-points evolve into 8 pairs of Weyl-nodes out of which 4 are within the first BZ and the rest are outside it. (iii) Mn$_3$Sn and Mn$_3$Ge, in which nodal-ring is reported to be observed in the absence of SOC\cite{yang2017topological1}. Also, this work shows that SOC gaps these nodal rings but leaves discrete band-touching points which are found to be Weyl-nodes. However, this work does not highlight the number of nodal-loop found in these materials. Furthermore, the information regarding the number of Weyl-nodes which evolve from each nodal-loop is also missing. (iv) TaAs class of WSMs, which are reported to show 4 gapless nodal rings in the mirror plane\cite{lv2015observation,weng2015weyl}. Furthermore, these works highlight that under the effect of SOC, each nodal-rings evolve into 3 pairs of Weyl-nodes. 

Above discussions clearly show the existence of a nodal-ring due to accidental degeneracy. In the presence of SOC, this nodal-ring is evolving into sets of Weyl-nodes. This nodal-ring and associated Weyl-nodes can be considered as arising from fundamental degeneracy as mentioned earlier. The rest of the nodal-rings and associated Weyl-nodes can be generated by applying symmetry operations and hence can be considered as derived ones. Above study did not look at the possibility of the existence of sets of \textbf{\textit{k}}-points in the nodal-line forming an arc as a fundamental entity, if the application of space-group symmetry operations including the TR symmetry in the reciprocal space on this arc generates full nodal-ring as well as other nodal-rings. Such arc in this work is termed as \textit{nodal-arc}. The possibility of the existence of such \textit{nodal-arc} will help in understanding the role of accidental degeneracy in the formation of number of Weyl-nodes in a given material in a better way. Furthermore, the energy gap created at these \textit{nodal-arcs} in the presence of SOC is also not well explored. Investigating this aspect will be helpful in understanding the evolution Weyl-nodes from these \textit{nodal-arcs}. Moreover, the information about the orbitals involved in the band-inversion in these materials is also lacking. Furthermore in case of HfC, it is reported that beyond a critical strength of SOC, the material changes its phase to topological insulator\cite{yu2017nodal}. Such a study of the materials response to the strength of SOC is missing for other mentioned compounds. Also, study of sensitivity of nodal-ring to the change in hybridization-strength due to the variation in lattice parameters is missing for the above mentioned materials. Such study will help in understanding the effect of pressure and strain on nodal-line and Weyl-nodes.

In this work we have made an attempt to address the above mentioned issues related to the formation of nodal-rings and their evolution into Weyl-nodes in the presence of SOC in TaAs \& TaP materials. Any such attempts require the accurate finding of band-touching points in terms of their coordinates and energy. To ensure this, here we have used \textit{PY-Nodes} code which is efficient in finding nodes using the \textit{first-principle} approach\cite{pandey2022py}. This study clearly establishes out that each nodal-ring is composed of two arcs out of which one is the \textit{nodal-arc} and other can be obtained from symmetry operations. In the presence of SOC, the \textit{nodal-arc} is evolving into two pairs of Weyl-nodes out of which one is situated closer to the middle of the \textit{nodal-arc} and the other one at the end which is also shared with the other arc forming the nodal-ring. This causes the evolution of each nodal-ring into three pairs of Weyl-nodes.


\section{Computational details}
 The density functional theory (DFT) based calculations are carried out using WIEN2k package\cite{wien2k}. PBESol, which is based on generalised gradient approximation (GGA), is used as an exchange-correlation functional in these computations\cite{perdew2008restoring}. Both the compounds crystallize in body-centered tetragonal crystal structure with the space group $I4_1md$\cite{grassano2018validity}. The lattice constants for TaAs\cite{TaAs} (TaP\cite{TaP1,TaP2}) are \textit{a}=3.4824 \AA\hspace{0.03in} \& \textit{c}=11.8038 \AA \hspace{0.03in}(\textit{a}=3.3300 \AA\hspace{0.03in} \& \textit{c}= 11.3900 \AA). The Wyckoff position of Ta is (0,0,0) while for As \& P, it is (0,0,u) with u=0.4176 (0.4160) for As (P)\cite{TaAs,TaP1,TaP2}. For both TaAs and TaP, the self-consistent ground state energy calculations are performed over a \textit{k}-mesh size of 10$\times$10$\times$10 with the charge convergence limit set to $10^{-10}$ Rydberg. Using these ground state structures, the nodes-finding calculations are carried out via \textit{PY-Nodes} code\cite{pandey2022py}. This code is based on \textit{Nelder-Mead's function minimization} method. The code is very efficient in searching the accurate coordinates of nodes using the \textit{first-principle} approach.

\section{Results and Discussion}

\subsection{Formation of \textit{nodal-arc}}

First of all, we would like to investigate the possibility of the existence of \textit{nodal-arc} arising as a result of fundamental degeneracy. For this, a detailed analysis of the formation of nodal-rings will be helpful. In this direction, firstly the orbitals leading to bands-inversion is explored. For this, the orbitals contributing to the topmost valence band (TVB) and bottommost conduction band (BCB) of TaAs \& TaP along the high symmetric directions \textit{$\Gamma$-N-Z-P-X-$\Gamma$-P-N} is investigated. It is found that these bands are mostly contributed by Ta-5\textit{d} and As-4\textit{p}/P-3\textit{p} orbitals. At the atomic level, the energy of Ta-5\textit{d} is higher than the energy of As-4\textit{p} and P-3\textit{p} orbitals. This generally suggests that the TVB must have higher contribution from As-4\textit{p} and P-3\textit{p} while BCB must have higher contribution from Ta-5\textit{d}. The results obtained from calculation of the character contribution to these bands is shown in fig. 1. It is seen that in all the region, except a small one in the $N-Z$ direction, the expected trend as discussed above is found to be followed. Within this small region along $N-Z$ direction, the orbital contribution is inverted where TVB has higher contribution from Ta-5\textit{d} while BCB has higher contribution from As-4\textit{p} and P-3\textit{p} orbitals. Thus, band-inversion is found to occur in this region. In this work, such a character inversion within a band is referred as \textit{intraband dp inversion (IDPI)}. Such kind of band-inversion generally favours the occurrence of non-trivial band-crossings. Also, as the character variation across the given high-symmetric path is found to be exactly similar in TaAs \& TaP, identical bands-crossings is expected to occur in both the materials. The calculations is carried out to search for the bands-crossing. For both the materials, four nodal-rings situated on the mirror planes defined by $k_x$=0 and $k_y$=0 planes are obtained. In the further investigation, these nodal-rings are found to be situated around the $N-Z$ direction. Hence, exploring if the \textit{IDPI} is solely responsible for the formation of these rings is required. In this regard, the character contribution to TVB \& BCB is studied within the plane containing the nodal-ring. It is found that \textit{IDPI} is occurring for the \textbf{\textit{k}}-points within the ring when compared with those outside it. This clearly establishes that \textit{IDPI} is leading to the formation of these nodal-rings.

\begin{figure}
    \centering
    {
        \includegraphics[width=0.80\linewidth, height=4.45cm]{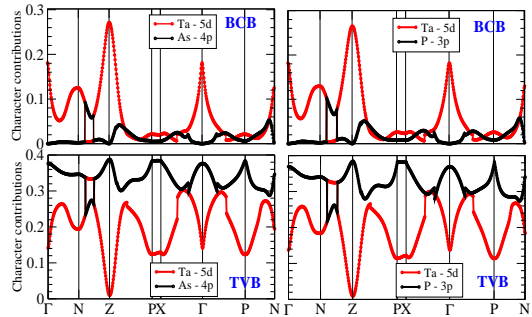}
    }
    \caption
    {{ {(Color online) \textit{IDPI} in \textit{N-Z} direction for TVB \& BCB corresponding to TaAs \& TaP.} }
    }
    \label{fig:foobar}
\end{figure}
\begin{figure}
    \centering
    \subfigure[]
    {
        \includegraphics[width=0.55\linewidth, height=3.5cm]{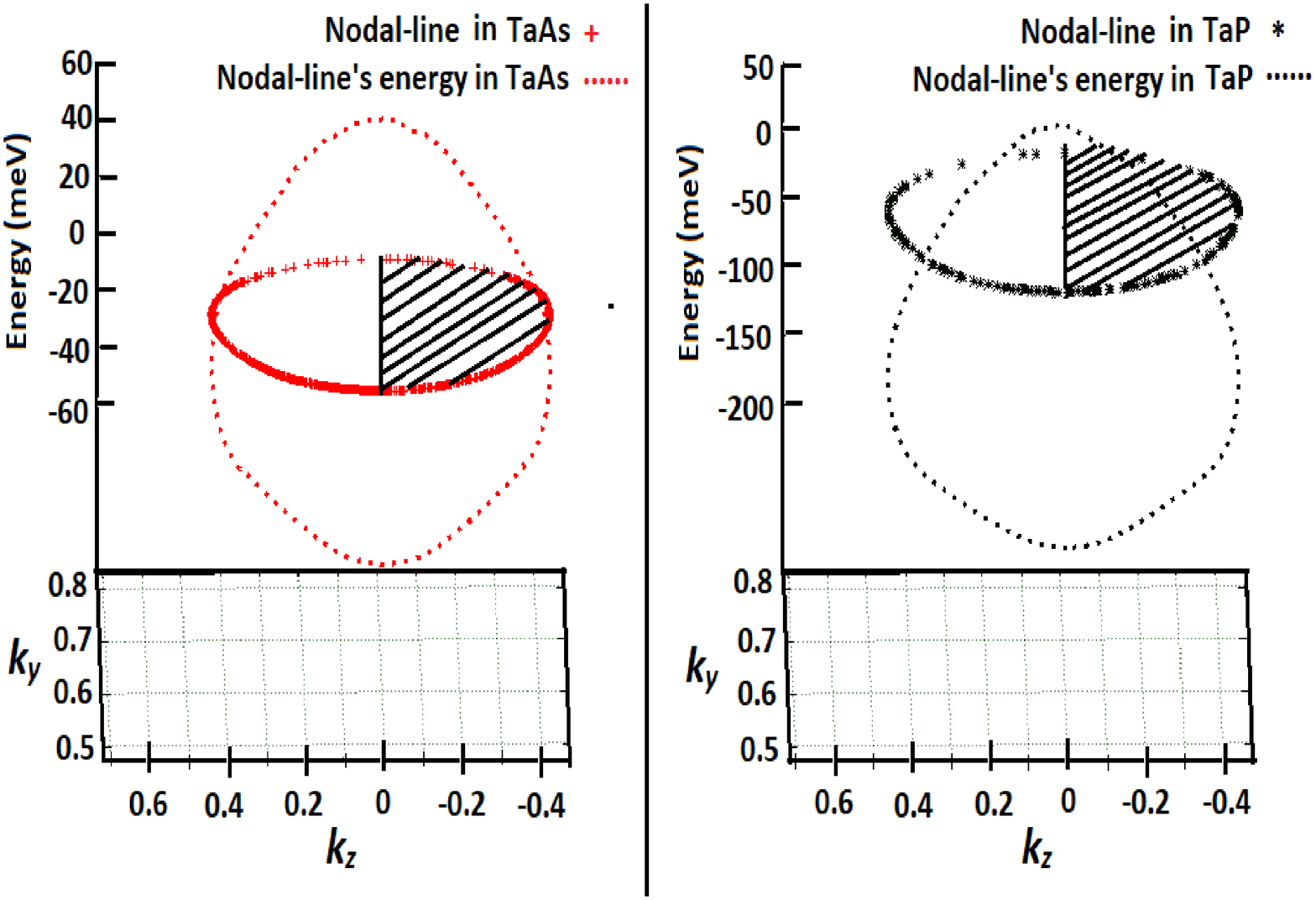}
    }
    \subfigure[]
    {
        \includegraphics[width=0.35\linewidth, height=3.3cm]{fig2b.eps}
    }
    \caption
    { {{ (Color online) Plot (a) shows the energy of nodal-ring in TaAs \& TaP, respectively. The obtained energy is symmetric across $k_z$=0 plane. Plot (b) shows the quantitative comparison of energy of \textit{nodal-arc} along $k_y$ axis for TaAs \& TaP.}}
    }
    \label{fig:foobar}
\end{figure}

Now we are interested in exploring the features of nodal-rings in further details. The four nodal-rings in these materials are related with space-group symmetry operations. Thus, our analysis is concentrated over a single ring. The energy of the nodal-rings in TaAs \& TaP is calculated and the obtained results are shown in fig. 2(a). It is found that these rings are not situated at a constant energy but are spread withing a particular energy range. The total energy spread in case of TaP ($\sim$204 meV) is obtained to be nearly double than than in the case of TaAs ($\sim$109 meV). Apart from this, the obtained energy of these rings is symmetric with respect to $k_z$=0 plane. Thus, an extra symmetry, resulting from the combined effect of mirror symmetry and the TR symmetry, is there within the nodal-ring. This prevents us to consider the ring as coming from fundamental degeneracy. However, portion of the ring surrounding the shaded region does not involve any kind of symmetry operations. So, it can be considered as \textit{nodal-arc} which is arising due to fundamental degeneracy. This \textit{nodal-arc} in TaAs is found to be much more spread on the energy scale than in case of TaP. The quantitative comparison of energy spread of \textit{nodal-arc} in TaAs \& TaP is presented in fig. 2(b). As the \textit{nodal-arc} is found to be the fundamental-degeneracy in these materials, it must be capable of generating all the rings upon the application of symmetry operations associated with TaAs \& TaP. This is discussed further.

Figure 3 schematically shows the step-by-step generation of other nodal-lines degeneracy from the \textit{nodal-arc} upon the application of symmetry operations associated with these materials. The symmetry operations associated with the primitive unit-cell of these materials includes- (i) 4-fold rotation ($C_4$) symmetry across $z$=0 axis, (ii) mirror symmetry about $x$=0 plane, (iii) mirror symmetry about $y$=0 plane and (iv) time-reversal (TR) symmetry. These symmetry operations results in the set of symmetries associated with the \textbf{\textit{k}}-space which includes- (i) $C_4$ symmetry across $k_z$=0 axis, (ii) mirror symmetry about $k_x$=0 plane, (iii) mirror symmetry about $k_y$=0 plane and (iv) degeneracy of states at \textbf{\textit{k}} and \textbf{-\textit{k}} due to TR symmetry\cite{bradley1972mathematical}. Now let us apply these operations on \textit{nodal-arc} to find out other degenerate regions in the BZ. Firstly, application of $C_4$ symmetry on the \textit{nodal-arc} in $k_x$=0 plane generates another similar arc in $k_y$=0 plane which is shown in fig. 3 (b). Nextly, the application of mirror symmetry about $k_y$=0 plane produces another similar arc in the $k_x$=0 plane but in the negative direction of $k_y$ axis. This is shown in fig. 3(c). This operation has no effect on the blue arc situated on $k_y$=0 plane since it is lying on the mirror plane. Similarly, the application of mirror symmetry operation across $k_x$=0 plane generates a blue arc on $k_y$=0 plane but in the negative direction of $k_x$ axis. The resultant set of arcs produced due to these operations is shown in fig. 3(d). It is seen from the figure that these arcs are situated in the negative direction of $k_z$ axis. Finally, considering the \textbf{\textit{k}} and \textbf{-\textit{k}} points degeneracy due to TR symmetry results in the generation of four other arcs in the positive $k_z$ axis. All these arcs together form four nodal-rings as shown in fig. 3 (e). The four nodal-lines generated in this way is found to be same as those obtained from nodes calculations. This suggest the existence of \textit{nodal-arc} as basic units for the creation of all the four rings found in these materials.

\begin{figure}
    \centering
    {
        \includegraphics[width=0.95\linewidth, height=5.0cm]{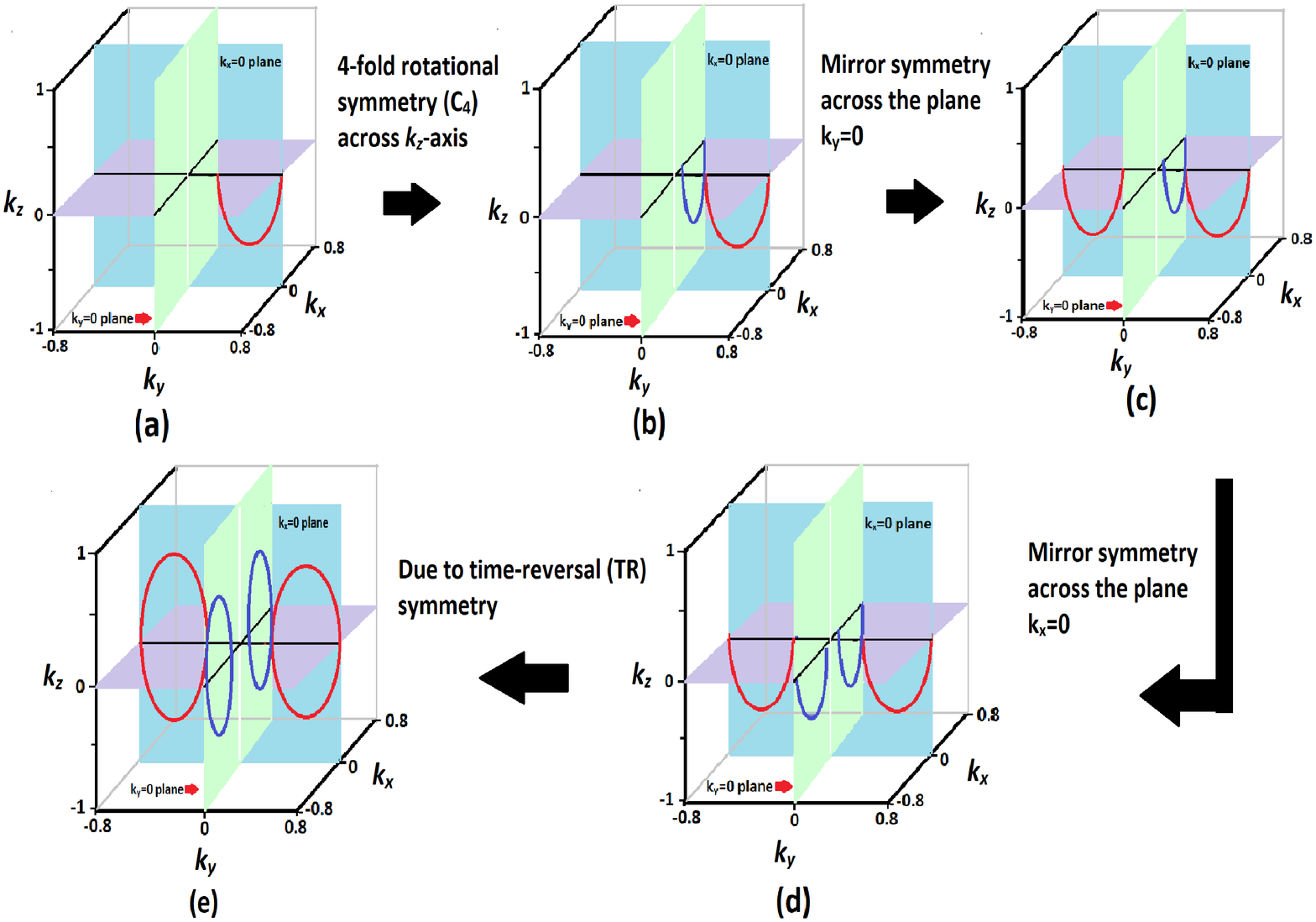}
    }
    \caption
    { {{(Color online) Schematic diagram showing the evolution of nodal-rings from \textit{nodal-arc} upon the application of space-group symmetry operations including TR symmetry in TaAs \& TaP.}}
    }
    \label{fig:foobar}
\end{figure}

\begin{figure}
    \centering
    {
        \includegraphics[width=0.80\linewidth, height=4.0cm]{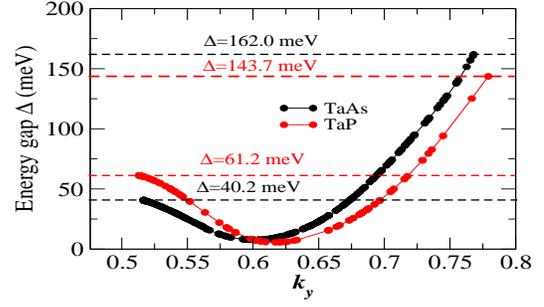}
    }  
    \caption
    { {{(Color online) Energy gap created at the \textit{nodal-arc} due to SOC in TaAs \& TaP.}}
    }
    \label{fig:foobar}
\end{figure}

\begin{figure}
    \centering
    {
        \includegraphics[width=0.95\linewidth, height=3.5cm]{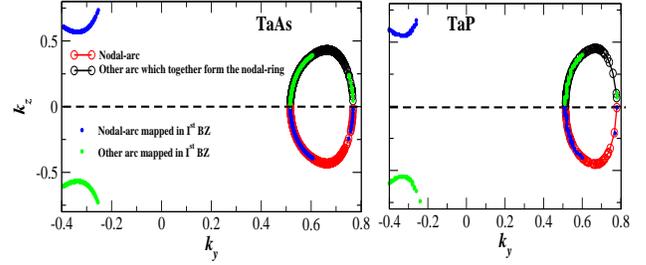}
    }
    \caption
    { {{(Color online) Plot showing the portion of \textit{nodal-arcs} associated with the first Brillouin zone of TaAs \& TaP.}}
    }
    \label{fig:foobar}
\end{figure}

\begin{table*}\label{tabc}
\caption{\label{tabc}
{The coordinates and energy of the representative Weyl-nodes (W1 \& W2) in TaAs \& TaP obtained corresponding to different strength of SOC used in the calculations.}}
 \centering
 \begin{tabular}{|p{1.5cm}|c|c|c|c|c|c|c|}
   \hline
   \multirow{2}{1cm}{{\textbf{Material}}} &{\textbf{Value of \textit{c}}} & {\textbf{SOC-strength per Ta atom}}& \multicolumn{2}{c|}{\textbf{W1 point}}& \multicolumn{2}{c|}{\textbf{W2 point}}  \\
   \cline{4-7}
  & \textbf{(atomic unit)} & \textbf{(meV)} & \textbf{  Coordinates  } & \textbf{Energy (meV)} & \textbf{  Coordinates  } & \textbf{Energy (meV)}  \\
   \hline
          & 137.03     & 754.8      & (0.51,0.01,0.00)     & -20.1  & (0.28,0.02,0.59) & -19.3 \\
  TaAs    & 130.18     & 2939.6     & (0.52,0.01,0.00)     & -9.6   & (0.28,0.02,0.58) & -21.4 \\
          & 123.32     & 3395.9     & (0.51,0.01,0.00)     &  3.2   & (0.28,0.03,0.57) & -29.8 \\
          &            &             &                      &         &                  &        \\
          & 137.03     & 718.8      & (0.51,0.01,0.00)     & -58.7  & (0.27,0.02,0.59) &  6.6   \\
   TaP    & 130.18     & 3041.0     & (0.51,0.01,0.00)     & -58.5  & (0.27,0.02,0.58) & -3.8   \\
          & 123.32     & 3512.9     & (0.51,0.01,0.00)     & -61.4  & (0.27,0.02,0.57) & -23.2  \\
   \hline
 \end{tabular}
 
\end{table*}

It is important to note here that the fundamental degeneracy associated with the \textit{nodal-arc} can be related to the fundamental domain of group theory. Suppose $P_n$ denotes the topological space and $G$ represents a group. Let $Y$ denotes the discrete subgroup of $G$. Then a fundamental domain for $P_n$/$Y$ is a subset of $P_n$ such that the union of the images under the action of $Y$ covers $P_n$, i.e.,\cite{pjm}
\begin{equation}
 \bigcup_{y\epsilon Y}y(P_n/Y) =P_n
\end{equation}
In the present case, the symmetry operations related with the nodal-rings can be considered to form a group $G$. Thus, the elements of $G$ will be \{$I$, $C_4$, $M_x$, $M_y$, $M_xM_y$, $C_4M_y$, $C_4M_x$, $C_4M_xM_y$, $T$, $TC_4$, $TM_x$, $TM_y$, $TM_xM_y$, $TC_4M_y$, $TC_4M_x$, $TC_4M_xM_y$\}. Here, $I$, $C_4$, $M_x$, $M_y$ \& $T$ stand for identity operation, 4-fold rotation about $k_z$=0 axis, mirror symmetry about $k_x$=0 plane, mirror symmetry about $k_y$=0 plane and TR symmetry, respectively.
Now, according to the above definition of fundamental domain, elements of $Y$ can be considered as: \{$I$, $C_4$, $M_x$, $C_4M_y$, $T$, $TC_4$, $TM_x$, $TC_4M_y$\}. Applying each operations of $Y$ on the \textit{nodal-arc} will give a unique arc in different region of the BZ. Arcs obtained from all the eight operations of $Y$ subset of $G$ will together form the four nodal-rings. Thus, \textit{nodal-arc} is the fundamental domain for the $Y$ subset as defined above. 

\subsection{Evolution of \textit{nodal-arc} to Weyl-nodes}

Analysis of the energy gap created at the \textit{nodal-arc} due to SOC may be a helpful step towards exploring its evolution into Weyl-nodes. The obtained results in this regard for TaAs \& TaP are shown in fig. 4. The gap created at each point of the arc is found to be not of uniform magnitude. This indicates that the SOC-strength in these materials is $\textbf{\textit{k}}$-dependent. However, the energy gap along these \textit{nodal-arcs} is obtained to be systematic in nature. Along the increasing direction of $k_y$, the gap is found to first decrease to attain a minimum non-zero value. This value corresponds to 7.7 (5.6) meV in TaAs (TaP). On further increase in the value of $k_y$, the energy gap again starts rising. The maximum value of the gap obtained in TaAs (TaP) is found to be 162.0 (143.7) meV. Before moving further, it is important to highlight here that these \textit{nodal-arcs} are found to be extended beyond the boundaries of the first BZ. The results regarding this is shown in fig. 5. Here, the red arc denotes the \textit{nodal-arc} which along with the black arc completes the nodal-ring. Furthermore, when the red and black arcs are mapped to first BZ, the blue and green colored arcs are obtained, respectively. While major portions of these new arcs are situated at the same points as the red \& black arcs, some parts of it are found in entirely new region as indicated in the figure. This verifies the extension of \textit{nodal-arcs} beyond the first BZ. It is important to mention here that there are  inconsistency in the results regarding the spatial extent of nodal-rings of these materials in the literature. Several works predict that these rings are not extended beyond the boundaries of first BZ\cite{huang2015weyl,lee2015fermi,belopolski2016criteria}. On the other hand, some other results indicate that such loops extend beyond the first BZ\cite{sun2016strong,fang2016topological}. The findings in the present work regarding \textit{nodal-arc} ensures that the nodal-rings in these materials are extended beyond the boundaries of first BZ. Also, it must be noted here that the term \textit{nodal-arc} in further discussions will refer to regions obtained after mapping the previously defined \textit{nodal-arc} to the first BZ. Now moving further, although SOC destroys the \textit{nodal-arc} degeneracy in these materials, it is found to create four new degenerate points close to the plane containing the \textit{nodal-arc}. These points correspond to two pairs of Weyl-nodes. The coordinates of Weyl-nodes evolving from the \textit{nodal-arc} and the other arc which together form a nodal-ring is shown in fig. 6. The two nodes forming a Weyl pair are found to be situated on either sides of the \textit{nodal-arc} and also perpendicular to the plane containing the arc. Furthermore, one pair of Weyl-nodes is situated closer to the middle of the \textit{nodal-arc} and the other one is obtained at the end as shown in the figure. In this way, one \textit{nodal-arc} is found to evolve into two pairs of Weyl-nodes. However, the pair which is situated closer to the end of the \textit{nodal-arc} has the shared contribution from both the arc forming the ring. This is because it is situated closer to the joining-point of both the arcs forming the ring, which can be seen from the figure. Thus, both the arcs forming the ring are found to collectively evolve into 3 pairs of Weyl-nodes. This discussion provides a justification for the evolution of each nodal-ring into three pairs of Weyl-nodes in these WSMs. The next point of discussion here is whether the strength of SOC affects the number of Weyl-nodes into which these arcs evolve. This is discussed further.

\begin{figure}
    \centering
    {
        \includegraphics[width=0.80\linewidth, height=4.5cm]{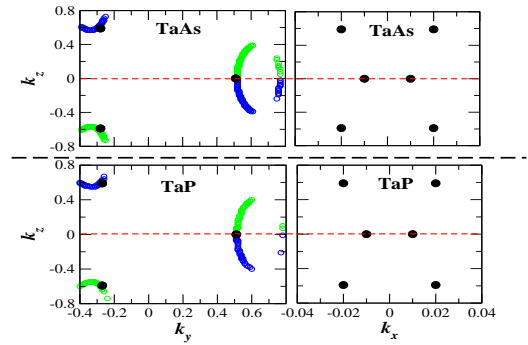}
    }  
    \caption
    { {{(Color online) The position of Weyl-nodes in TaAs \& TaP with respect to the coordinates of \textit{nodal-arcs} from which they are evolved due to SOC.}}
    }
    \label{fig:foobar}
\end{figure}

The SOC Hamiltonian ($H_{so}$) is inversely proportional to $c^2$, where $c$ is the velocity of light\cite{2001wien2k}. In the present work, the value of $c$ is changed to vary the strength of SOC in the calculations. The ground state energy calculations in the presence of SOC are performed using three different values of $c$: 137.03, 130.18  \& 123.32. These values are in atomic units. For both the materials, the strength of SOC obtained corresponding to these values of $c$ are mentioned in table I. It is to be noted here that SOC-strength is calculated by finding the difference between the total energy per Ta atom obtained before and after the inclusion of SOC. It is seen that with the decrease in the value of $c$ from $\sim$137 to $\sim$123, the strength of SOC in TaAs (TaP) increases from $\sim$0.7 eV ($\sim$0.7 eV) to $\sim$3.4 eV ($\sim$3.5 eV). The evolution of \textit{nodal-arc} into Weyl-nodes under the effect of these three different strength of SOC is studied for these materials. Corresponding to these strength of SOC, the \textit{nodal-arcs} in TaAs \& TaP are found to evolve into only two pairs of Weyl-nodes. Thus, the given change in SOC-strength has no effect on the number of Weyl-nodes into which the arcs are evolving. The coordinates and energy of one node from each pair are mentioned in table I. Since both the points forming the Weyl pair is at equidistant from the plane containing the \textit{nodal-arc}, only one is mentioned in the table. Also, the Weyl-nodes, which is situated closer to the end of the \textit{nodal-arc} is indicated as W1 point while the one closer to the middle of the arc is indicated as W2 point in the table. It is found that with the increase in the strength of SOC, the coordinates of the Weyl-nodes remained same in both the WSMs. However, the energy of W1 and W2 points changes considerably. As can be seen from the table, with the given increase in strength of SOC, the W1 \& W2 points of TaAs (TaP) are getting far-from (closer-to) each other in terms of their energy. For instance, with the increase in SOC-strength from $\sim$0.7 eV to $\sim$3.0 eV, the energy gap between W1-W2 points gets changed from 0.77 (65.32) meV to 11.79 (54.68) meV in case of TaAs (TaP).

As mentioned in the introduction, HfC, a WSM candidate, shows phase transition from WSM to topological insulator when strength of SOC is increased beyond a certain limit ($\sim$ 1 eV)\cite{yu2017nodal}. However, similar behaviour is not observed in case of TaAs \& TaP. The existence of Weyl-nodes in these materials are found to be robust against much stronger SOC-strength, more than 3 times, as compared to that for HfC. It is found that Weyl-nodes in TaAs (TaP) remain intact at their position for the increase in the SOC-strength from $\sim$ 0.75 ($\sim$ 0.72) eV to $\sim$ 3.39 ($\sim$ 3.51) eV. These results highlight that within the range of $\sim$ 0.7-3.5 eV, SOC-strength is acting at perturbative level on the top of non-SOC Hamiltonian of these materials. Thus, this strength of SOC may not be enough to create the gap at the Weyl-nodes.

\setlength{\parindent}{3em}
\setlength{\parskip}{0.2em}

\begin{figure}
    \centering
    {
        \includegraphics[width=0.80\linewidth, height=4.0cm]{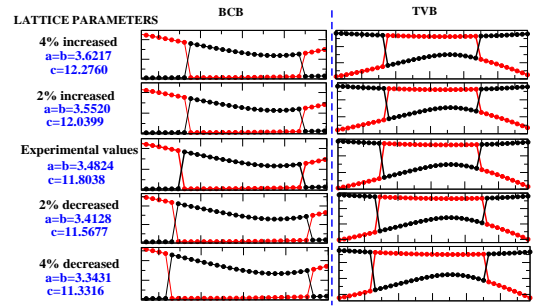}
    }  
    \caption
    { {{(Color online) Effect of change in lattice parameters on \textit{IDPI} in TaAs.}}
    }
    \label{fig:foobar}
\end{figure}

\subsection{Effect of hybridization-strength on \textit{IDPI} and nodal-rings}

The change in lattice parameters affects the inter-atomic distances and hence the hybridization-strength of atomic orbitals. This will certainly affect the character contributions to the states associated with any given bands. As the \textit{IDPI} is closely related to the character contributions to the TVB \& BCB, it becomes necessary to find-out if \textit{IDPI} is sensitive to change in lattice parameters. In this direction, the calculations of character contributions of Ta-5\textit{d} \& As-4\textit{p} to TVB \& BCB is carried out for TaAs by varying the lattice parameters from -4\% to 4\% in the step of 2\% from the experimental value. Corresponding to all these sets, the band-character calculations show that \textit{IDPI} occurs along \textit{N-Z} direction in TVB and BCB. The obtained results are shown in fig. 7. It is seen from the figure that with the increase in the lattice parameters, the spatial extent of the \textit{IDPI}, in \textit{N-Z} direction, decreases. Similar results have been also obtained for TaP. These observations clearly indicate that extent of \textit{IDPI} in TaAs and TaP is extremely sensitive to the change in lattice parameters. It is already discovered that the \textit{nodal-arc} and hence the nodal-rings in these materials is resulting from \textit{IDPI}. Thus, above discussion suggests that the size of nodal-rings in these materials must reduce with the increase in the value of lattice parameters. In this direction, further investigation is carried out and the obtained results are discussed below.
\begin{figure}
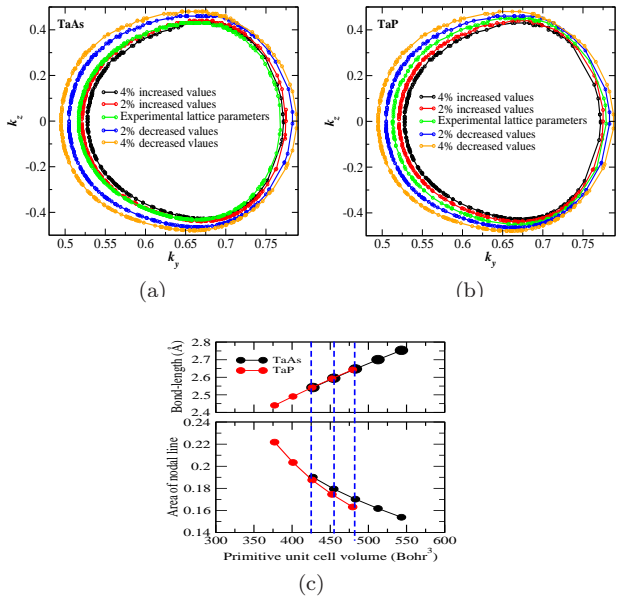

    \centering
        \subfigure[]
    {
        \includegraphics[width=0.45\linewidth, height=3.5cm]{fig8a.eps}
    }
    \subfigure[]
    {
        \includegraphics[width=0.45\linewidth, height=3.5cm]{fig8b.eps}
    }\vspace*{0.1in}
    \subfigure[]
    {
        \includegraphics[width=0.45\linewidth, height=3.0cm]{fig8c.eps}
    }
    \caption
    { {{(Color online) Plots (a) \& (b) shows the change in nodal-ring in $k_x$=0 plane obtained corresponding to change in lattice parameters of TaAs \& TaP, respectively. (c) Area of nodal-ring versus primitive unit cell volume for TaAs \& TaP.}}
    }
    \label{fig:foobar}
\end{figure}
For the above mentioned values of lattice parameters, nodal-ring calculations are performed for TaAs \& TaP. For both the materials, one of the nodal-rings obtained in $k_x$=0 plane for different values of lattice parameters are compared. Corresponding plots for TaAs \& TaP are shown in fig. 8(a) and 8(b), respectively. It is seen from these figures that with the increase in the lattice parameters, the size of nodal-ring decreases for both the materials. Thus, the results obtained are found to be consistent with the prediction made above. Furthermore, it is obvious that the change in lattice parameters will result in the change in the volume of the primitive unit cell and the Ta-As/Ta-P bond-length. The variation of the area of nodal-line and the bond-length with the change in volume of the primitive unit cell, for both the materials, is shown in fig. 8(c). It is seen from the figure that with the increase in primitive unit cell volume of both the materials, the bond-length of Ta-As \& Ta-P linearly increases whereas the area of nodal-line linearly decreases. The rate of decrease in the area of nodal-line with the increase in bond-length is found to be more in TaP than in TaAs. With the increase in bond-length from 2.59 to 2.64 \AA, the area of nodal-ring in TaAs decreases from 0.179 to 0.170. Here, the area is unitless since the values on $k_y$ and $k_z$ axis denote the coefficients of reciprocal lattice vectors and thus have no units. Moving further, for the same change in the bond-length in TaP, the area of nodal-ring decreases from 0.174 to 0.163. Thus, with the same increase in the bond-length, increase in the hybridization-strength in TaP is comparatively higher than TaAs. These observations indicate the role of hybridization of atomic orbitals in the occurrence of \textit{IDPI} and the formation of nodal-rings. This is explicitly discussed further.

\begin{figure}
    \centering
    {
        \includegraphics[width=0.65\linewidth, height=5.5cm]{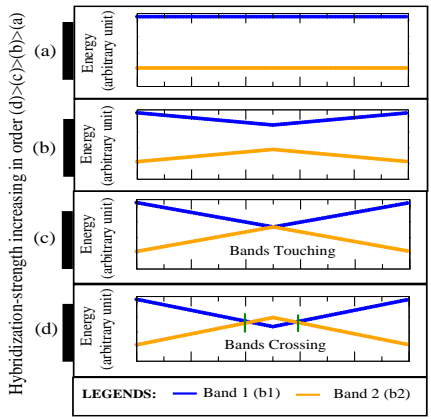}
    }
    \caption
    { {{(Color online) Schematic showing the role of hybridization in the formation of nodal-ring.}}
    }
    \label{fig:foobar}
\end{figure}

As already discussed, TVB \& BCB in TaAs (TaP) are mostly contributed by Ta-5\textit{d} and As-4\textit{p} (P-3\textit{p}) orbitals. When atoms are very far from each other, BCB (b1) will have only contributions from Ta-5\textit{d} while the TVB (b2) will be contributed by As-4\textit{p}/P-3\textit{p}. In that case, these bands will be non-dispersive as shown in fig. 9 (a). When the atoms come closer, their atomic orbitals will overlap resulting in the hybridization. This will make these bands dispersive and bring them closer to each other as shown in fig. 9 (b). With the further decrease in the inter-atomic distances, the overlapping and hence hybridization-strength will increase and the bands will approach each other to a greater extent to form a touching point or node. This is shown in fig. 9 (c). In similar manner, any further increase in the hybridization-strength will result in the crossing of these bands to form a nodal-ring, which can be seen in fig. 9 (d). It is intuitive from the figure that with further rise in hybridization-strength, the size of nodal-ring will increase. In this way, hybridization of atomic orbitals leads to the formation of nodal-rings. This also explains why decrease in the bond-length is resulting into the increase in the area of nodal-rings.

\section{Conclusions} 

In this work, several aspects regarding nodal-lines and its evolution into Weyl-nodes in TaAs \& TaP are explored using the \textit{first-principle} approach. These includes- (i) orbitals responsible for bands-inversion, (ii) possibility of the existence of \textit{nodal-arc} which generates full nodal-ring and other rings upon the application of symmetry operations, (iii) energy spread of such \textit{nodal-arcs}, (iv) effect of SOC on these arcs, (v) number of Weyl-nodes evolving from these \textit{nodal-arcs} under the effect of SOC, (vi) effect of change in SOC-strength on the Weyl-nodes and (vi) sensitivity of the nodal-rings to the change in hybridization-strength. The orbitals leading to band-inversion are obtained as Ta-5\textit{d} \& As-4\textit{p} (P-3\textit{p}). The nodal-ring is found to be formed from a \textit{nodal-arc} and another arc which is related to the \textit{nodal-arc} through the combination of mirror-symmetry and TR symmetry. The arc in TaAs (TaP) is found to be spread within $\sim$ 109 ($\sim$ 204) meV around the Fermi-level. These \textit{nodal-arcs} get fully gaped under the effect of SOC. The maximum value of the gap obtained in TaAs (TaP) is found to be 161.95 (143.66) while the minimum value corresponds to 7.68 (5.58) meV. SOC leads to the evolution of \textit{nodal-arc} into 2 pairs of Weyl-nodes whose coordinates are found to be robust to the change in SOC-strength from $\sim$ 0.7-3.5 eV. The nodal-rings are found to be highly sensitive of the change in hybridization-strength. Increase in bond-length from 2.59 to 2.64 \AA \hspace{0.01in} reduces the area of nodal-ring from 0.179 (0.174) to 0.170 (0.163) in TaAs (TaP). The findings of this work make the picture more clear regarding the role of accidental degeneracy in the formation of nodal-rings and the Weyl-nodes in TaAs \& TaP.

\section{References}
\bibliography{MS_revised}

\begin{thebibliography}{44}%
\makeatletter
\providecommand \@ifxundefined [1]{%
 \@ifx{#1\undefined}
}%
\providecommand \@ifnum [1]{%
 \ifnum #1\expandafter \@firstoftwo
 \else \expandafter \@secondoftwo
 \fi
}%
\providecommand \@ifx [1]{%
 \ifx #1\expandafter \@firstoftwo
 \else \expandafter \@secondoftwo
 \fi
}%
\providecommand \natexlab [1]{#1}%
\providecommand \enquote  [1]{``#1''}%
\providecommand \bibnamefont  [1]{#1}%
\providecommand \bibfnamefont [1]{#1}%
\providecommand \citenamefont [1]{#1}%
\providecommand \href@noop [0]{\@secondoftwo}%
\providecommand \href [0]{\begingroup \@sanitize@url \@href}%
\providecommand \@href[1]{\@@startlink{#1}\@@href}%
\providecommand \@@href[1]{\endgroup#1\@@endlink}%
\providecommand \@sanitize@url [0]{\catcode `\\12\catcode `\$12\catcode
  `\&12\catcode `\#12\catcode `\^12\catcode `\_12\catcode `\%12\relax}%
\providecommand \@@startlink[1]{}%
\providecommand \@@endlink[0]{}%
\providecommand \url  [0]{\begingroup\@sanitize@url \@url }%
\providecommand \@url [1]{\endgroup\@href {#1}{\urlprefix }}%
\providecommand \urlprefix  [0]{URL }%
\providecommand \Eprint [0]{\href }%
\providecommand \doibase [0]{http://dx.doi.org/}%
\providecommand \selectlanguage [0]{\@gobble}%
\providecommand \bibinfo  [0]{\@secondoftwo}%
\providecommand \bibfield  [0]{\@secondoftwo}%
\providecommand \translation [1]{[#1]}%
\providecommand \BibitemOpen [0]{}%
\providecommand \bibitemStop [0]{}%
\providecommand \bibitemNoStop [0]{.\EOS\space}%
\providecommand \EOS [0]{\spacefactor3000\relax}%
\providecommand \BibitemShut  [1]{\csname bibitem#1\endcsname}%
\let\auto@bib@innerbib\@empty
\bibitem [{\citenamefont {Weng}\ \emph
  {et~al.}(2015{\natexlab{a}})\citenamefont {Weng}, \citenamefont {Fang},
  \citenamefont {Fang}, \citenamefont {Bernevig},\ and\ \citenamefont
  {Dai}}]{weng2015weyl}%
  \BibitemOpen
  \bibfield  {author} {\bibinfo {author} {\bibfnamefont {H.}~\bibnamefont
  {Weng}}, \bibinfo {author} {\bibfnamefont {C.}~\bibnamefont {Fang}}, \bibinfo
  {author} {\bibfnamefont {Z.}~\bibnamefont {Fang}}, \bibinfo {author}
  {\bibfnamefont {B.~A.}\ \bibnamefont {Bernevig}}, \ and\ \bibinfo {author}
  {\bibfnamefont {X.}~\bibnamefont {Dai}},\ }\href@noop {} {\bibfield
  {journal} {\bibinfo  {journal} {Physical Review X}\ }\textbf {\bibinfo
  {volume} {5}},\ \bibinfo {pages} {011029} (\bibinfo {year}
  {2015}{\natexlab{a}})}\BibitemShut {NoStop}%
\bibitem [{\citenamefont {Soluyanov}\ \emph {et~al.}(2015)\citenamefont
  {Soluyanov}, \citenamefont {Gresch}, \citenamefont {Wang}, \citenamefont
  {Wu}, \citenamefont {Troyer}, \citenamefont {Dai},\ and\ \citenamefont
  {Bernevig}}]{soluyanov2015type}%
  \BibitemOpen
  \bibfield  {author} {\bibinfo {author} {\bibfnamefont {A.~A.}\ \bibnamefont
  {Soluyanov}}, \bibinfo {author} {\bibfnamefont {D.}~\bibnamefont {Gresch}},
  \bibinfo {author} {\bibfnamefont {Z.}~\bibnamefont {Wang}}, \bibinfo {author}
  {\bibfnamefont {Q.}~\bibnamefont {Wu}}, \bibinfo {author} {\bibfnamefont
  {M.}~\bibnamefont {Troyer}}, \bibinfo {author} {\bibfnamefont
  {X.}~\bibnamefont {Dai}}, \ and\ \bibinfo {author} {\bibfnamefont {B.~A.}\
  \bibnamefont {Bernevig}},\ }\href@noop {} {\bibfield  {journal} {\bibinfo
  {journal} {Nature}\ }\textbf {\bibinfo {volume} {527}},\ \bibinfo {pages}
  {495} (\bibinfo {year} {2015})}\BibitemShut {NoStop}%
\bibitem [{\citenamefont {Lv}\ \emph {et~al.}(2015{\natexlab{a}})\citenamefont
  {Lv}, \citenamefont {Weng}, \citenamefont {Fu}, \citenamefont {Wang},
  \citenamefont {Miao}, \citenamefont {Ma}, \citenamefont {Richard},
  \citenamefont {Huang}, \citenamefont {Zhao}, \citenamefont {Chen} \emph
  {et~al.}}]{lv2015experimental}%
  \BibitemOpen
  \bibfield  {author} {\bibinfo {author} {\bibfnamefont {B.}~\bibnamefont
  {Lv}}, \bibinfo {author} {\bibfnamefont {H.}~\bibnamefont {Weng}}, \bibinfo
  {author} {\bibfnamefont {B.}~\bibnamefont {Fu}}, \bibinfo {author}
  {\bibfnamefont {X.~P.}\ \bibnamefont {Wang}}, \bibinfo {author}
  {\bibfnamefont {H.}~\bibnamefont {Miao}}, \bibinfo {author} {\bibfnamefont
  {J.}~\bibnamefont {Ma}}, \bibinfo {author} {\bibfnamefont {P.}~\bibnamefont
  {Richard}}, \bibinfo {author} {\bibfnamefont {X.}~\bibnamefont {Huang}},
  \bibinfo {author} {\bibfnamefont {L.}~\bibnamefont {Zhao}}, \bibinfo {author}
  {\bibfnamefont {G.}~\bibnamefont {Chen}},  \emph {et~al.},\ }\href@noop {}
  {\bibfield  {journal} {\bibinfo  {journal} {Physical Review X}\ }\textbf
  {\bibinfo {volume} {5}},\ \bibinfo {pages} {031013} (\bibinfo {year}
  {2015}{\natexlab{a}})}\BibitemShut {NoStop}%
\bibitem [{\citenamefont {Pandey}\ \emph {et~al.}(2021)\citenamefont {Pandey},
  \citenamefont {Sihi},\ and\ \citenamefont {Pandey}}]{pandey2021anab}%
  \BibitemOpen
  \bibfield  {author} {\bibinfo {author} {\bibfnamefont {V.}~\bibnamefont
  {Pandey}}, \bibinfo {author} {\bibfnamefont {A.}~\bibnamefont {Sihi}}, \ and\
  \bibinfo {author} {\bibfnamefont {S.~K.}\ \bibnamefont {Pandey}},\
  }\href@noop {} {\bibfield  {journal} {\bibinfo  {journal} {Journal of
  Physics: Condensed Matter}\ }\textbf {\bibinfo {volume} {33}},\ \bibinfo
  {pages} {475503} (\bibinfo {year} {2021})}\BibitemShut {NoStop}%
\bibitem [{\citenamefont {Neupane}\ \emph {et~al.}(2014)\citenamefont
  {Neupane}, \citenamefont {Xu}, \citenamefont {Sankar}, \citenamefont
  {Alidoust}, \citenamefont {Bian}, \citenamefont {Liu}, \citenamefont
  {Belopolski}, \citenamefont {Chang}, \citenamefont {Jeng}, \citenamefont
  {Lin} \emph {et~al.}}]{neupane2014observation}%
  \BibitemOpen
  \bibfield  {author} {\bibinfo {author} {\bibfnamefont {M.}~\bibnamefont
  {Neupane}}, \bibinfo {author} {\bibfnamefont {S.-Y.}\ \bibnamefont {Xu}},
  \bibinfo {author} {\bibfnamefont {R.}~\bibnamefont {Sankar}}, \bibinfo
  {author} {\bibfnamefont {N.}~\bibnamefont {Alidoust}}, \bibinfo {author}
  {\bibfnamefont {G.}~\bibnamefont {Bian}}, \bibinfo {author} {\bibfnamefont
  {C.}~\bibnamefont {Liu}}, \bibinfo {author} {\bibfnamefont {I.}~\bibnamefont
  {Belopolski}}, \bibinfo {author} {\bibfnamefont {T.-R.}\ \bibnamefont
  {Chang}}, \bibinfo {author} {\bibfnamefont {H.-T.}\ \bibnamefont {Jeng}},
  \bibinfo {author} {\bibfnamefont {H.}~\bibnamefont {Lin}},  \emph {et~al.},\
  }\href@noop {} {\bibfield  {journal} {\bibinfo  {journal} {Nature
  communications}\ }\textbf {\bibinfo {volume} {5}},\ \bibinfo {pages} {3786}
  (\bibinfo {year} {2014})}\BibitemShut {NoStop}%
\bibitem [{\citenamefont {Steinberg}\ \emph {et~al.}(2014)\citenamefont
  {Steinberg}, \citenamefont {Young}, \citenamefont {Zaheer}, \citenamefont
  {Kane}, \citenamefont {Mele},\ and\ \citenamefont
  {Rappe}}]{steinberg2014bulk}%
  \BibitemOpen
  \bibfield  {author} {\bibinfo {author} {\bibfnamefont {J.~A.}\ \bibnamefont
  {Steinberg}}, \bibinfo {author} {\bibfnamefont {S.~M.}\ \bibnamefont
  {Young}}, \bibinfo {author} {\bibfnamefont {S.}~\bibnamefont {Zaheer}},
  \bibinfo {author} {\bibfnamefont {C.}~\bibnamefont {Kane}}, \bibinfo {author}
  {\bibfnamefont {E.}~\bibnamefont {Mele}}, \ and\ \bibinfo {author}
  {\bibfnamefont {A.~M.}\ \bibnamefont {Rappe}},\ }\href@noop {} {\bibfield
  {journal} {\bibinfo  {journal} {Physical review letters}\ }\textbf {\bibinfo
  {volume} {112}},\ \bibinfo {pages} {036403} (\bibinfo {year}
  {2014})}\BibitemShut {NoStop}%
\bibitem [{\citenamefont {Yang}\ and\ \citenamefont
  {Nagaosa}(2014)}]{yang2014classification}%
  \BibitemOpen
  \bibfield  {author} {\bibinfo {author} {\bibfnamefont {B.-J.}\ \bibnamefont
  {Yang}}\ and\ \bibinfo {author} {\bibfnamefont {N.}~\bibnamefont {Nagaosa}},\
  }\href@noop {} {\bibfield  {journal} {\bibinfo  {journal} {Nature
  communications}\ }\textbf {\bibinfo {volume} {5}},\ \bibinfo {pages} {4898}
  (\bibinfo {year} {2014})}\BibitemShut {NoStop}%
\bibitem [{\citenamefont {Burkov}\ \emph {et~al.}(2011)\citenamefont {Burkov},
  \citenamefont {Hook},\ and\ \citenamefont {Balents}}]{burkov2011topological}%
  \BibitemOpen
  \bibfield  {author} {\bibinfo {author} {\bibfnamefont {A.}~\bibnamefont
  {Burkov}}, \bibinfo {author} {\bibfnamefont {M.}~\bibnamefont {Hook}}, \ and\
  \bibinfo {author} {\bibfnamefont {L.}~\bibnamefont {Balents}},\ }\href@noop
  {} {\bibfield  {journal} {\bibinfo  {journal} {Physical Review B}\ }\textbf
  {\bibinfo {volume} {84}},\ \bibinfo {pages} {235126} (\bibinfo {year}
  {2011})}\BibitemShut {NoStop}%
\bibitem [{\citenamefont {Yu}\ \emph {et~al.}(2015)\citenamefont {Yu},
  \citenamefont {Weng}, \citenamefont {Fang}, \citenamefont {Dai},\ and\
  \citenamefont {Hu}}]{yu2015topological}%
  \BibitemOpen
  \bibfield  {author} {\bibinfo {author} {\bibfnamefont {R.}~\bibnamefont
  {Yu}}, \bibinfo {author} {\bibfnamefont {H.}~\bibnamefont {Weng}}, \bibinfo
  {author} {\bibfnamefont {Z.}~\bibnamefont {Fang}}, \bibinfo {author}
  {\bibfnamefont {X.}~\bibnamefont {Dai}}, \ and\ \bibinfo {author}
  {\bibfnamefont {X.}~\bibnamefont {Hu}},\ }\href@noop {} {\bibfield  {journal}
  {\bibinfo  {journal} {Physical review letters}\ }\textbf {\bibinfo {volume}
  {115}},\ \bibinfo {pages} {036807} (\bibinfo {year} {2015})}\BibitemShut
  {NoStop}%
\bibitem [{\citenamefont {Xu}\ \emph {et~al.}(2017)\citenamefont {Xu},
  \citenamefont {Yu}, \citenamefont {Fang}, \citenamefont {Dai},\ and\
  \citenamefont {Weng}}]{xu2017topological}%
  \BibitemOpen
  \bibfield  {author} {\bibinfo {author} {\bibfnamefont {Q.}~\bibnamefont
  {Xu}}, \bibinfo {author} {\bibfnamefont {R.}~\bibnamefont {Yu}}, \bibinfo
  {author} {\bibfnamefont {Z.}~\bibnamefont {Fang}}, \bibinfo {author}
  {\bibfnamefont {X.}~\bibnamefont {Dai}}, \ and\ \bibinfo {author}
  {\bibfnamefont {H.}~\bibnamefont {Weng}},\ }\href@noop {} {\bibfield
  {journal} {\bibinfo  {journal} {Physical Review B}\ }\textbf {\bibinfo
  {volume} {95}},\ \bibinfo {pages} {045136} (\bibinfo {year}
  {2017})}\BibitemShut {NoStop}%
\bibitem [{\citenamefont {Belopolski}\ \emph {et~al.}(2016)\citenamefont
  {Belopolski}, \citenamefont {Xu}, \citenamefont {Sanchez}, \citenamefont
  {Chang}, \citenamefont {Guo}, \citenamefont {Neupane}, \citenamefont {Zheng},
  \citenamefont {Lee}, \citenamefont {Huang}, \citenamefont {Bian} \emph
  {et~al.}}]{belopolski2016criteria}%
  \BibitemOpen
  \bibfield  {author} {\bibinfo {author} {\bibfnamefont {I.}~\bibnamefont
  {Belopolski}}, \bibinfo {author} {\bibfnamefont {S.-Y.}\ \bibnamefont {Xu}},
  \bibinfo {author} {\bibfnamefont {D.~S.}\ \bibnamefont {Sanchez}}, \bibinfo
  {author} {\bibfnamefont {G.}~\bibnamefont {Chang}}, \bibinfo {author}
  {\bibfnamefont {C.}~\bibnamefont {Guo}}, \bibinfo {author} {\bibfnamefont
  {M.}~\bibnamefont {Neupane}}, \bibinfo {author} {\bibfnamefont
  {H.}~\bibnamefont {Zheng}}, \bibinfo {author} {\bibfnamefont {C.-C.}\
  \bibnamefont {Lee}}, \bibinfo {author} {\bibfnamefont {S.-M.}\ \bibnamefont
  {Huang}}, \bibinfo {author} {\bibfnamefont {G.}~\bibnamefont {Bian}},  \emph
  {et~al.},\ }\href@noop {} {\bibfield  {journal} {\bibinfo  {journal}
  {Physical review letters}\ }\textbf {\bibinfo {volume} {116}},\ \bibinfo
  {pages} {066802} (\bibinfo {year} {2016})}\BibitemShut {NoStop}%
\bibitem [{\citenamefont {Hosur}(2012)}]{hosur2012friedel}%
  \BibitemOpen
  \bibfield  {author} {\bibinfo {author} {\bibfnamefont {P.}~\bibnamefont
  {Hosur}},\ }\href@noop {} {\bibfield  {journal} {\bibinfo  {journal}
  {Physical Review B}\ }\textbf {\bibinfo {volume} {86}},\ \bibinfo {pages}
  {195102} (\bibinfo {year} {2012})}\BibitemShut {NoStop}%
\bibitem [{\citenamefont {Sihi}\ and\ \citenamefont
  {Pandey}(2022)}]{sihi2022evidence}%
  \BibitemOpen
  \bibfield  {author} {\bibinfo {author} {\bibfnamefont {A.}~\bibnamefont
  {Sihi}}\ and\ \bibinfo {author} {\bibfnamefont {S.~K.}\ \bibnamefont
  {Pandey}},\ }\href@noop {} {\bibfield  {journal} {\bibinfo  {journal}
  {Journal of Physics: Condensed Matter}\ }\textbf {\bibinfo {volume} {34}},\
  \bibinfo {pages} {325601} (\bibinfo {year} {2022})}\BibitemShut {NoStop}%
\bibitem [{\citenamefont {Miao}\ \emph {et~al.}(2018)\citenamefont {Miao},
  \citenamefont {Zhang}, \citenamefont {Wang}, \citenamefont {Meyers},
  \citenamefont {Said}, \citenamefont {Wang}, \citenamefont {Shi},
  \citenamefont {Weng}, \citenamefont {Fang},\ and\ \citenamefont
  {Dean}}]{miao2018observation}%
  \BibitemOpen
  \bibfield  {author} {\bibinfo {author} {\bibfnamefont {H.}~\bibnamefont
  {Miao}}, \bibinfo {author} {\bibfnamefont {T.}~\bibnamefont {Zhang}},
  \bibinfo {author} {\bibfnamefont {L.}~\bibnamefont {Wang}}, \bibinfo {author}
  {\bibfnamefont {D.}~\bibnamefont {Meyers}}, \bibinfo {author} {\bibfnamefont
  {A.}~\bibnamefont {Said}}, \bibinfo {author} {\bibfnamefont {Y.}~\bibnamefont
  {Wang}}, \bibinfo {author} {\bibfnamefont {Y.}~\bibnamefont {Shi}}, \bibinfo
  {author} {\bibfnamefont {H.}~\bibnamefont {Weng}}, \bibinfo {author}
  {\bibfnamefont {Z.}~\bibnamefont {Fang}}, \ and\ \bibinfo {author}
  {\bibfnamefont {M.}~\bibnamefont {Dean}},\ }\href@noop {} {\bibfield
  {journal} {\bibinfo  {journal} {Physical review letters}\ }\textbf {\bibinfo
  {volume} {121}},\ \bibinfo {pages} {035302} (\bibinfo {year}
  {2018})}\BibitemShut {NoStop}%
\bibitem [{\citenamefont {Armitage}\ \emph {et~al.}(2018)\citenamefont
  {Armitage}, \citenamefont {Mele},\ and\ \citenamefont
  {Vishwanath}}]{RevModPhys.90.015001}%
  \BibitemOpen
  \bibfield  {author} {\bibinfo {author} {\bibfnamefont {N.~P.}\ \bibnamefont
  {Armitage}}, \bibinfo {author} {\bibfnamefont {E.~J.}\ \bibnamefont {Mele}},
  \ and\ \bibinfo {author} {\bibfnamefont {A.}~\bibnamefont {Vishwanath}},\
  }\href {\doibase 10.1103/RevModPhys.90.015001} {\bibfield  {journal}
  {\bibinfo  {journal} {Rev. Mod. Phys.}\ }\textbf {\bibinfo {volume} {90}},\
  \bibinfo {pages} {015001} (\bibinfo {year} {2018})}\BibitemShut {NoStop}%
\bibitem [{\citenamefont {Quan}\ \emph {et~al.}(2017)\citenamefont {Quan},
  \citenamefont {Yin},\ and\ \citenamefont {Pickett}}]{quan2017single}%
  \BibitemOpen
  \bibfield  {author} {\bibinfo {author} {\bibfnamefont {Y.}~\bibnamefont
  {Quan}}, \bibinfo {author} {\bibfnamefont {Z.}~\bibnamefont {Yin}}, \ and\
  \bibinfo {author} {\bibfnamefont {W.}~\bibnamefont {Pickett}},\ }\href@noop
  {} {\bibfield  {journal} {\bibinfo  {journal} {Physical review letters}\
  }\textbf {\bibinfo {volume} {118}},\ \bibinfo {pages} {176402} (\bibinfo
  {year} {2017})}\BibitemShut {NoStop}%
\bibitem [{\citenamefont {Wu}\ \emph {et~al.}(2016)\citenamefont {Wu},
  \citenamefont {Wang}, \citenamefont {Mun}, \citenamefont {Johnson},
  \citenamefont {Mou}, \citenamefont {Huang}, \citenamefont {Lee},
  \citenamefont {Bud’ko}, \citenamefont {Canfield},\ and\ \citenamefont
  {Kaminski}}]{wu2016dirac}%
  \BibitemOpen
  \bibfield  {author} {\bibinfo {author} {\bibfnamefont {Y.}~\bibnamefont
  {Wu}}, \bibinfo {author} {\bibfnamefont {L.-L.}\ \bibnamefont {Wang}},
  \bibinfo {author} {\bibfnamefont {E.}~\bibnamefont {Mun}}, \bibinfo {author}
  {\bibfnamefont {D.~D.}\ \bibnamefont {Johnson}}, \bibinfo {author}
  {\bibfnamefont {D.}~\bibnamefont {Mou}}, \bibinfo {author} {\bibfnamefont
  {L.}~\bibnamefont {Huang}}, \bibinfo {author} {\bibfnamefont
  {Y.}~\bibnamefont {Lee}}, \bibinfo {author} {\bibfnamefont {S.~L.}\
  \bibnamefont {Bud’ko}}, \bibinfo {author} {\bibfnamefont {P.~C.}\
  \bibnamefont {Canfield}}, \ and\ \bibinfo {author} {\bibfnamefont
  {A.}~\bibnamefont {Kaminski}},\ }\href@noop {} {\bibfield  {journal}
  {\bibinfo  {journal} {Nature Physics}\ }\textbf {\bibinfo {volume} {12}},\
  \bibinfo {pages} {667} (\bibinfo {year} {2016})}\BibitemShut {NoStop}%
\bibitem [{\citenamefont {Kim}\ \emph {et~al.}(2015)\citenamefont {Kim},
  \citenamefont {Wieder}, \citenamefont {Kane},\ and\ \citenamefont
  {Rappe}}]{kim2015dirac}%
  \BibitemOpen
  \bibfield  {author} {\bibinfo {author} {\bibfnamefont {Y.}~\bibnamefont
  {Kim}}, \bibinfo {author} {\bibfnamefont {B.~J.}\ \bibnamefont {Wieder}},
  \bibinfo {author} {\bibfnamefont {C.}~\bibnamefont {Kane}}, \ and\ \bibinfo
  {author} {\bibfnamefont {A.~M.}\ \bibnamefont {Rappe}},\ }\href@noop {}
  {\bibfield  {journal} {\bibinfo  {journal} {Physical review letters}\
  }\textbf {\bibinfo {volume} {115}},\ \bibinfo {pages} {036806} (\bibinfo
  {year} {2015})}\BibitemShut {NoStop}%
\bibitem [{\citenamefont {Yang}\ \emph {et~al.}(2014)\citenamefont {Yang},
  \citenamefont {Pan},\ and\ \citenamefont {Zhang}}]{yang2014dirac}%
  \BibitemOpen
  \bibfield  {author} {\bibinfo {author} {\bibfnamefont {S.~A.}\ \bibnamefont
  {Yang}}, \bibinfo {author} {\bibfnamefont {H.}~\bibnamefont {Pan}}, \ and\
  \bibinfo {author} {\bibfnamefont {F.}~\bibnamefont {Zhang}},\ }\href@noop {}
  {\bibfield  {journal} {\bibinfo  {journal} {Physical review letters}\
  }\textbf {\bibinfo {volume} {113}},\ \bibinfo {pages} {046401} (\bibinfo
  {year} {2014})}\BibitemShut {NoStop}%
\bibitem [{\citenamefont {Chiu}\ and\ \citenamefont
  {Schnyder}(2014)}]{chiu2014classification}%
  \BibitemOpen
  \bibfield  {author} {\bibinfo {author} {\bibfnamefont {C.-K.}\ \bibnamefont
  {Chiu}}\ and\ \bibinfo {author} {\bibfnamefont {A.~P.}\ \bibnamefont
  {Schnyder}},\ }\href@noop {} {\bibfield  {journal} {\bibinfo  {journal}
  {Physical Review B}\ }\textbf {\bibinfo {volume} {90}},\ \bibinfo {pages}
  {205136} (\bibinfo {year} {2014})}\BibitemShut {NoStop}%
\bibitem [{\citenamefont {Bian}\ \emph {et~al.}(2016)\citenamefont {Bian},
  \citenamefont {Chang}, \citenamefont {Sankar}, \citenamefont {Xu},
  \citenamefont {Zheng}, \citenamefont {Neupert}, \citenamefont {Chiu},
  \citenamefont {Huang}, \citenamefont {Chang}, \citenamefont {Belopolski}
  \emph {et~al.}}]{bian2016topological}%
  \BibitemOpen
  \bibfield  {author} {\bibinfo {author} {\bibfnamefont {G.}~\bibnamefont
  {Bian}}, \bibinfo {author} {\bibfnamefont {T.-R.}\ \bibnamefont {Chang}},
  \bibinfo {author} {\bibfnamefont {R.}~\bibnamefont {Sankar}}, \bibinfo
  {author} {\bibfnamefont {S.-Y.}\ \bibnamefont {Xu}}, \bibinfo {author}
  {\bibfnamefont {H.}~\bibnamefont {Zheng}}, \bibinfo {author} {\bibfnamefont
  {T.}~\bibnamefont {Neupert}}, \bibinfo {author} {\bibfnamefont {C.-K.}\
  \bibnamefont {Chiu}}, \bibinfo {author} {\bibfnamefont {S.-M.}\ \bibnamefont
  {Huang}}, \bibinfo {author} {\bibfnamefont {G.}~\bibnamefont {Chang}},
  \bibinfo {author} {\bibfnamefont {I.}~\bibnamefont {Belopolski}},  \emph
  {et~al.},\ }\href@noop {} {\bibfield  {journal} {\bibinfo  {journal} {Nature
  communications}\ }\textbf {\bibinfo {volume} {7}},\ \bibinfo {pages} {10556}
  (\bibinfo {year} {2016})}\BibitemShut {NoStop}%
\bibitem [{\citenamefont {Huang}\ \emph {et~al.}(2015)\citenamefont {Huang},
  \citenamefont {Xu}, \citenamefont {Belopolski}, \citenamefont {Lee},
  \citenamefont {Chang}, \citenamefont {Wang}, \citenamefont {Alidoust},
  \citenamefont {Bian}, \citenamefont {Neupane}, \citenamefont {Zhang} \emph
  {et~al.}}]{huang2015weyl}%
  \BibitemOpen
  \bibfield  {author} {\bibinfo {author} {\bibfnamefont {S.-M.}\ \bibnamefont
  {Huang}}, \bibinfo {author} {\bibfnamefont {S.-Y.}\ \bibnamefont {Xu}},
  \bibinfo {author} {\bibfnamefont {I.}~\bibnamefont {Belopolski}}, \bibinfo
  {author} {\bibfnamefont {C.-C.}\ \bibnamefont {Lee}}, \bibinfo {author}
  {\bibfnamefont {G.}~\bibnamefont {Chang}}, \bibinfo {author} {\bibfnamefont
  {B.}~\bibnamefont {Wang}}, \bibinfo {author} {\bibfnamefont {N.}~\bibnamefont
  {Alidoust}}, \bibinfo {author} {\bibfnamefont {G.}~\bibnamefont {Bian}},
  \bibinfo {author} {\bibfnamefont {M.}~\bibnamefont {Neupane}}, \bibinfo
  {author} {\bibfnamefont {C.}~\bibnamefont {Zhang}},  \emph {et~al.},\
  }\href@noop {} {\bibfield  {journal} {\bibinfo  {journal} {Nature
  communications}\ }\textbf {\bibinfo {volume} {6}},\ \bibinfo {pages} {1}
  (\bibinfo {year} {2015})}\BibitemShut {NoStop}%
\bibitem [{\citenamefont {Tang}\ \emph {et~al.}(2016)\citenamefont {Tang},
  \citenamefont {Zhou}, \citenamefont {Xu},\ and\ \citenamefont
  {Zhang}}]{tang2016dirac}%
  \BibitemOpen
  \bibfield  {author} {\bibinfo {author} {\bibfnamefont {P.}~\bibnamefont
  {Tang}}, \bibinfo {author} {\bibfnamefont {Q.}~\bibnamefont {Zhou}}, \bibinfo
  {author} {\bibfnamefont {G.}~\bibnamefont {Xu}}, \ and\ \bibinfo {author}
  {\bibfnamefont {S.-C.}\ \bibnamefont {Zhang}},\ }\href@noop {} {\bibfield
  {journal} {\bibinfo  {journal} {Nature Physics}\ }\textbf {\bibinfo {volume}
  {12}},\ \bibinfo {pages} {1100} (\bibinfo {year} {2016})}\BibitemShut
  {NoStop}%
\bibitem [{\citenamefont {Chen}\ \emph {et~al.}(2017)\citenamefont {Chen},
  \citenamefont {Xu}, \citenamefont {Jiang}, \citenamefont {Wu}, \citenamefont
  {Qi}, \citenamefont {Yang}, \citenamefont {Wang}, \citenamefont {Sun},
  \citenamefont {Schr{\"o}ter}, \citenamefont {Yang} \emph
  {et~al.}}]{chen2017dirac}%
  \BibitemOpen
  \bibfield  {author} {\bibinfo {author} {\bibfnamefont {C.}~\bibnamefont
  {Chen}}, \bibinfo {author} {\bibfnamefont {X.}~\bibnamefont {Xu}}, \bibinfo
  {author} {\bibfnamefont {J.}~\bibnamefont {Jiang}}, \bibinfo {author}
  {\bibfnamefont {S.-C.}\ \bibnamefont {Wu}}, \bibinfo {author} {\bibfnamefont
  {Y.}~\bibnamefont {Qi}}, \bibinfo {author} {\bibfnamefont {L.}~\bibnamefont
  {Yang}}, \bibinfo {author} {\bibfnamefont {M.}~\bibnamefont {Wang}}, \bibinfo
  {author} {\bibfnamefont {Y.}~\bibnamefont {Sun}}, \bibinfo {author}
  {\bibfnamefont {N.}~\bibnamefont {Schr{\"o}ter}}, \bibinfo {author}
  {\bibfnamefont {H.}~\bibnamefont {Yang}},  \emph {et~al.},\ }\href@noop {}
  {\bibfield  {journal} {\bibinfo  {journal} {Physical Review B}\ }\textbf
  {\bibinfo {volume} {95}},\ \bibinfo {pages} {125126} (\bibinfo {year}
  {2017})}\BibitemShut {NoStop}%
\bibitem [{\citenamefont {Chen}\ \emph {et~al.}(2015)\citenamefont {Chen},
  \citenamefont {Lu},\ and\ \citenamefont {Kee}}]{chen2015topological}%
  \BibitemOpen
  \bibfield  {author} {\bibinfo {author} {\bibfnamefont {Y.}~\bibnamefont
  {Chen}}, \bibinfo {author} {\bibfnamefont {Y.-M.}\ \bibnamefont {Lu}}, \ and\
  \bibinfo {author} {\bibfnamefont {H.-Y.}\ \bibnamefont {Kee}},\ }\href@noop
  {} {\bibfield  {journal} {\bibinfo  {journal} {Nature communications}\
  }\textbf {\bibinfo {volume} {6}},\ \bibinfo {pages} {6593} (\bibinfo {year}
  {2015})}\BibitemShut {NoStop}%
\bibitem [{\citenamefont {Bzdu{\v{s}}ek}\ \emph {et~al.}(2016)\citenamefont
  {Bzdu{\v{s}}ek}, \citenamefont {Wu}, \citenamefont {R{\"u}egg}, \citenamefont
  {Sigrist},\ and\ \citenamefont {Soluyanov}}]{bzduvsek2016nodal}%
  \BibitemOpen
  \bibfield  {author} {\bibinfo {author} {\bibfnamefont {T.}~\bibnamefont
  {Bzdu{\v{s}}ek}}, \bibinfo {author} {\bibfnamefont {Q.}~\bibnamefont {Wu}},
  \bibinfo {author} {\bibfnamefont {A.}~\bibnamefont {R{\"u}egg}}, \bibinfo
  {author} {\bibfnamefont {M.}~\bibnamefont {Sigrist}}, \ and\ \bibinfo
  {author} {\bibfnamefont {A.~A.}\ \bibnamefont {Soluyanov}},\ }\href@noop {}
  {\bibfield  {journal} {\bibinfo  {journal} {Nature}\ }\textbf {\bibinfo
  {volume} {538}},\ \bibinfo {pages} {75} (\bibinfo {year} {2016})}\BibitemShut
  {NoStop}%
\bibitem [{\citenamefont {Yang}\ \emph {et~al.}(2018)\citenamefont {Yang},
  \citenamefont {Moessner},\ and\ \citenamefont {Lim}}]{yang2018quantum}%
  \BibitemOpen
  \bibfield  {author} {\bibinfo {author} {\bibfnamefont {H.}~\bibnamefont
  {Yang}}, \bibinfo {author} {\bibfnamefont {R.}~\bibnamefont {Moessner}}, \
  and\ \bibinfo {author} {\bibfnamefont {L.-K.}\ \bibnamefont {Lim}},\
  }\href@noop {} {\bibfield  {journal} {\bibinfo  {journal} {Physical Review
  B}\ }\textbf {\bibinfo {volume} {97}},\ \bibinfo {pages} {165118} (\bibinfo
  {year} {2018})}\BibitemShut {NoStop}%
\bibitem [{\citenamefont {Liu}\ \emph {et~al.}(2018)\citenamefont {Liu},
  \citenamefont {Sun}, \citenamefont {Kumar}, \citenamefont {Muechler},
  \citenamefont {Sun}, \citenamefont {Jiao}, \citenamefont {Yang},
  \citenamefont {Liu}, \citenamefont {Liang}, \citenamefont {Xu} \emph
  {et~al.}}]{liu2018giant}%
  \BibitemOpen
  \bibfield  {author} {\bibinfo {author} {\bibfnamefont {E.}~\bibnamefont
  {Liu}}, \bibinfo {author} {\bibfnamefont {Y.}~\bibnamefont {Sun}}, \bibinfo
  {author} {\bibfnamefont {N.}~\bibnamefont {Kumar}}, \bibinfo {author}
  {\bibfnamefont {L.}~\bibnamefont {Muechler}}, \bibinfo {author}
  {\bibfnamefont {A.}~\bibnamefont {Sun}}, \bibinfo {author} {\bibfnamefont
  {L.}~\bibnamefont {Jiao}}, \bibinfo {author} {\bibfnamefont {S.-Y.}\
  \bibnamefont {Yang}}, \bibinfo {author} {\bibfnamefont {D.}~\bibnamefont
  {Liu}}, \bibinfo {author} {\bibfnamefont {A.}~\bibnamefont {Liang}}, \bibinfo
  {author} {\bibfnamefont {Q.}~\bibnamefont {Xu}},  \emph {et~al.},\
  }\href@noop {} {\bibfield  {journal} {\bibinfo  {journal} {Nature physics}\
  }\textbf {\bibinfo {volume} {14}},\ \bibinfo {pages} {1125} (\bibinfo {year}
  {2018})}\BibitemShut {NoStop}%
\bibitem [{\citenamefont {Yu}\ \emph {et~al.}(2017)\citenamefont {Yu},
  \citenamefont {Wu}, \citenamefont {Fang},\ and\ \citenamefont
  {Weng}}]{yu2017nodal}%
  \BibitemOpen
  \bibfield  {author} {\bibinfo {author} {\bibfnamefont {R.}~\bibnamefont
  {Yu}}, \bibinfo {author} {\bibfnamefont {Q.}~\bibnamefont {Wu}}, \bibinfo
  {author} {\bibfnamefont {Z.}~\bibnamefont {Fang}}, \ and\ \bibinfo {author}
  {\bibfnamefont {H.}~\bibnamefont {Weng}},\ }\href@noop {} {\bibfield
  {journal} {\bibinfo  {journal} {Physical review letters}\ }\textbf {\bibinfo
  {volume} {119}},\ \bibinfo {pages} {036401} (\bibinfo {year}
  {2017})}\BibitemShut {NoStop}%
\bibitem [{\citenamefont {Yang}\ \emph {et~al.}(2017)\citenamefont {Yang},
  \citenamefont {Sun}, \citenamefont {Zhang}, \citenamefont {Shi},
  \citenamefont {Parkin},\ and\ \citenamefont {Yan}}]{yang2017topological1}%
  \BibitemOpen
  \bibfield  {author} {\bibinfo {author} {\bibfnamefont {H.}~\bibnamefont
  {Yang}}, \bibinfo {author} {\bibfnamefont {Y.}~\bibnamefont {Sun}}, \bibinfo
  {author} {\bibfnamefont {Y.}~\bibnamefont {Zhang}}, \bibinfo {author}
  {\bibfnamefont {W.-J.}\ \bibnamefont {Shi}}, \bibinfo {author} {\bibfnamefont
  {S.~S.}\ \bibnamefont {Parkin}}, \ and\ \bibinfo {author} {\bibfnamefont
  {B.}~\bibnamefont {Yan}},\ }\href@noop {} {\bibfield  {journal} {\bibinfo
  {journal} {New Journal of Physics}\ }\textbf {\bibinfo {volume} {19}},\
  \bibinfo {pages} {015008} (\bibinfo {year} {2017})}\BibitemShut {NoStop}%
\bibitem [{\citenamefont {Lv}\ \emph {et~al.}(2015{\natexlab{b}})\citenamefont
  {Lv}, \citenamefont {Xu}, \citenamefont {Weng}, \citenamefont {Ma},
  \citenamefont {Richard}, \citenamefont {Huang}, \citenamefont {Zhao},
  \citenamefont {Chen}, \citenamefont {Matt}, \citenamefont {Bisti} \emph
  {et~al.}}]{lv2015observation}%
  \BibitemOpen
  \bibfield  {author} {\bibinfo {author} {\bibfnamefont {B.}~\bibnamefont
  {Lv}}, \bibinfo {author} {\bibfnamefont {N.}~\bibnamefont {Xu}}, \bibinfo
  {author} {\bibfnamefont {H.}~\bibnamefont {Weng}}, \bibinfo {author}
  {\bibfnamefont {J.}~\bibnamefont {Ma}}, \bibinfo {author} {\bibfnamefont
  {P.}~\bibnamefont {Richard}}, \bibinfo {author} {\bibfnamefont
  {X.}~\bibnamefont {Huang}}, \bibinfo {author} {\bibfnamefont
  {L.}~\bibnamefont {Zhao}}, \bibinfo {author} {\bibfnamefont {G.}~\bibnamefont
  {Chen}}, \bibinfo {author} {\bibfnamefont {C.}~\bibnamefont {Matt}}, \bibinfo
  {author} {\bibfnamefont {F.}~\bibnamefont {Bisti}},  \emph {et~al.},\
  }\href@noop {} {\bibfield  {journal} {\bibinfo  {journal} {Nature Physics}\
  }\textbf {\bibinfo {volume} {11}},\ \bibinfo {pages} {724} (\bibinfo {year}
  {2015}{\natexlab{b}})}\BibitemShut {NoStop}%
\bibitem [{\citenamefont {Pandey}\ and\ \citenamefont
  {Pandey}(2023)}]{pandey2022py}%
  \BibitemOpen
  \bibfield  {author} {\bibinfo {author} {\bibfnamefont {V.}~\bibnamefont
  {Pandey}}\ and\ \bibinfo {author} {\bibfnamefont {S.~K.}\ \bibnamefont
  {Pandey}},\ }\href {\doibase https://doi.org/10.1016/j.cpc.2022.108570}
  {\bibfield  {journal} {\bibinfo  {journal} {Computer Physics Communications}\
  }\textbf {\bibinfo {volume} {283}},\ \bibinfo {pages} {108570} (\bibinfo
  {year} {2023})}\BibitemShut {NoStop}%
\bibitem [{\citenamefont {Blaha}\ \emph {et~al.}(2020)\citenamefont {Blaha},
  \citenamefont {Schwarz}, \citenamefont {Tran}, \citenamefont {Laskowski},
  \citenamefont {Madsen},\ and\ \citenamefont {Marks}}]{wien2k}%
  \BibitemOpen
  \bibfield  {author} {\bibinfo {author} {\bibfnamefont {P.}~\bibnamefont
  {Blaha}}, \bibinfo {author} {\bibfnamefont {K.}~\bibnamefont {Schwarz}},
  \bibinfo {author} {\bibfnamefont {F.}~\bibnamefont {Tran}}, \bibinfo {author}
  {\bibfnamefont {R.}~\bibnamefont {Laskowski}}, \bibinfo {author}
  {\bibfnamefont {G.~K.}\ \bibnamefont {Madsen}}, \ and\ \bibinfo {author}
  {\bibfnamefont {L.~D.}\ \bibnamefont {Marks}},\ }\href@noop {} {\bibfield
  {journal} {\bibinfo  {journal} {The Journal of Chemical Physics}\ }\textbf
  {\bibinfo {volume} {152}},\ \bibinfo {pages} {074101} (\bibinfo {year}
  {2020})}\BibitemShut {NoStop}%
\bibitem [{\citenamefont {Perdew}\ \emph {et~al.}(2008)\citenamefont {Perdew},
  \citenamefont {Ruzsinszky}, \citenamefont {Csonka}, \citenamefont {Vydrov},
  \citenamefont {Scuseria}, \citenamefont {Constantin}, \citenamefont {Zhou},\
  and\ \citenamefont {Burke}}]{perdew2008restoring}%
  \BibitemOpen
  \bibfield  {author} {\bibinfo {author} {\bibfnamefont {J.~P.}\ \bibnamefont
  {Perdew}}, \bibinfo {author} {\bibfnamefont {A.}~\bibnamefont {Ruzsinszky}},
  \bibinfo {author} {\bibfnamefont {G.~I.}\ \bibnamefont {Csonka}}, \bibinfo
  {author} {\bibfnamefont {O.~A.}\ \bibnamefont {Vydrov}}, \bibinfo {author}
  {\bibfnamefont {G.~E.}\ \bibnamefont {Scuseria}}, \bibinfo {author}
  {\bibfnamefont {L.~A.}\ \bibnamefont {Constantin}}, \bibinfo {author}
  {\bibfnamefont {X.}~\bibnamefont {Zhou}}, \ and\ \bibinfo {author}
  {\bibfnamefont {K.}~\bibnamefont {Burke}},\ }\href@noop {} {\bibfield
  {journal} {\bibinfo  {journal} {Physical review letters}\ }\textbf {\bibinfo
  {volume} {100}},\ \bibinfo {pages} {136406} (\bibinfo {year}
  {2008})}\BibitemShut {NoStop}%
\bibitem [{\citenamefont {Grassano}\ \emph {et~al.}(2018)\citenamefont
  {Grassano}, \citenamefont {Pulci}, \citenamefont {Mosca~Conte},\ and\
  \citenamefont {Bechstedt}}]{grassano2018validity}%
  \BibitemOpen
  \bibfield  {author} {\bibinfo {author} {\bibfnamefont {D.}~\bibnamefont
  {Grassano}}, \bibinfo {author} {\bibfnamefont {O.}~\bibnamefont {Pulci}},
  \bibinfo {author} {\bibfnamefont {A.}~\bibnamefont {Mosca~Conte}}, \ and\
  \bibinfo {author} {\bibfnamefont {F.}~\bibnamefont {Bechstedt}},\ }\href@noop
  {} {\bibfield  {journal} {\bibinfo  {journal} {Scientific reports}\ }\textbf
  {\bibinfo {volume} {8}},\ \bibinfo {pages} {1} (\bibinfo {year}
  {2018})}\BibitemShut {NoStop}%
\bibitem [{\citenamefont {Furuseth}\ \emph {et~al.}(1965)\citenamefont
  {Furuseth}, \citenamefont {Selte}, \citenamefont {Kjekshus}, \citenamefont
  {Gronowitz}, \citenamefont {Hoffman},\ and\ \citenamefont
  {Westerdahl}}]{TaAs}%
  \BibitemOpen
  \bibfield  {author} {\bibinfo {author} {\bibfnamefont {S.}~\bibnamefont
  {Furuseth}}, \bibinfo {author} {\bibfnamefont {K.}~\bibnamefont {Selte}},
  \bibinfo {author} {\bibfnamefont {A.}~\bibnamefont {Kjekshus}}, \bibinfo
  {author} {\bibfnamefont {S.}~\bibnamefont {Gronowitz}}, \bibinfo {author}
  {\bibfnamefont {R.}~\bibnamefont {Hoffman}}, \ and\ \bibinfo {author}
  {\bibfnamefont {A.}~\bibnamefont {Westerdahl}},\ }\href@noop {} {\bibfield
  {journal} {\bibinfo  {journal} {Acta Chem. Scand}\ }\textbf {\bibinfo
  {volume} {19}},\ \bibinfo {pages} {42} (\bibinfo {year} {1965})}\BibitemShut
  {NoStop}%
\bibitem [{\citenamefont {Sapkota}\ \emph {et~al.}(2016)\citenamefont
  {Sapkota}, \citenamefont {Mukherjee},\ and\ \citenamefont {Mandrus}}]{TaP1}%
  \BibitemOpen
  \bibfield  {author} {\bibinfo {author} {\bibfnamefont {D.}~\bibnamefont
  {Sapkota}}, \bibinfo {author} {\bibfnamefont {R.}~\bibnamefont {Mukherjee}},
  \ and\ \bibinfo {author} {\bibfnamefont {D.}~\bibnamefont {Mandrus}},\
  }\href@noop {} {\bibfield  {journal} {\bibinfo  {journal} {Crystals}\
  }\textbf {\bibinfo {volume} {6}},\ \bibinfo {pages} {160} (\bibinfo {year}
  {2016})}\BibitemShut {NoStop}%
\bibitem [{\citenamefont {Weng}\ \emph
  {et~al.}(2015{\natexlab{b}})\citenamefont {Weng}, \citenamefont {Fang},
  \citenamefont {Fang}, \citenamefont {Bernevig},\ and\ \citenamefont
  {Dai}}]{TaP2}%
  \BibitemOpen
  \bibfield  {author} {\bibinfo {author} {\bibfnamefont {H.}~\bibnamefont
  {Weng}}, \bibinfo {author} {\bibfnamefont {C.}~\bibnamefont {Fang}}, \bibinfo
  {author} {\bibfnamefont {Z.}~\bibnamefont {Fang}}, \bibinfo {author}
  {\bibfnamefont {B.~A.}\ \bibnamefont {Bernevig}}, \ and\ \bibinfo {author}
  {\bibfnamefont {X.}~\bibnamefont {Dai}},\ }\href@noop {} {\bibfield
  {journal} {\bibinfo  {journal} {Physical Review X}\ }\textbf {\bibinfo
  {volume} {5}},\ \bibinfo {pages} {011029} (\bibinfo {year}
  {2015}{\natexlab{b}})}\BibitemShut {NoStop}%
\bibitem [{\citenamefont {Bradley}\ and\ \citenamefont
  {Cracknell}(1972)}]{bradley1972mathematical}%
  \BibitemOpen
  \bibfield  {author} {\bibinfo {author} {\bibfnamefont {C.}~\bibnamefont
  {Bradley}}\ and\ \bibinfo {author} {\bibfnamefont {A.}~\bibnamefont
  {Cracknell}},\ }\href {https://books.google.co.in/books?id=OKXvAAAAMAAJ}
  {\emph {\bibinfo {title} {The Mathematical Theory of Symmetry in Solids:
  Representation Theory for Point Groups and Space Groups}}}\ (\bibinfo
  {publisher} {Clarendon Press},\ \bibinfo {year} {1972})\BibitemShut {NoStop}%
\bibitem [{\citenamefont {Grenier}(1988)}]{pjm}%
  \BibitemOpen
  \bibfield  {author} {\bibinfo {author} {\bibfnamefont {D.}~\bibnamefont
  {Grenier}},\ }\href@noop {} {\bibfield  {journal} {\bibinfo  {journal}
  {Pacific Journal of Mathematics}\ }\textbf {\bibinfo {volume} {132}},\
  \bibinfo {pages} {293 } (\bibinfo {year} {1988})}\BibitemShut {NoStop}%
\bibitem [{\citenamefont {Lee}\ \emph {et~al.}(2015)\citenamefont {Lee},
  \citenamefont {Xu}, \citenamefont {Huang}, \citenamefont {Sanchez},
  \citenamefont {Belopolski}, \citenamefont {Chang}, \citenamefont {Bian},
  \citenamefont {Alidoust}, \citenamefont {Zheng}, \citenamefont {Neupane}
  \emph {et~al.}}]{lee2015fermi}%
  \BibitemOpen
  \bibfield  {author} {\bibinfo {author} {\bibfnamefont {C.-C.}\ \bibnamefont
  {Lee}}, \bibinfo {author} {\bibfnamefont {S.-Y.}\ \bibnamefont {Xu}},
  \bibinfo {author} {\bibfnamefont {S.-M.}\ \bibnamefont {Huang}}, \bibinfo
  {author} {\bibfnamefont {D.~S.}\ \bibnamefont {Sanchez}}, \bibinfo {author}
  {\bibfnamefont {I.}~\bibnamefont {Belopolski}}, \bibinfo {author}
  {\bibfnamefont {G.}~\bibnamefont {Chang}}, \bibinfo {author} {\bibfnamefont
  {G.}~\bibnamefont {Bian}}, \bibinfo {author} {\bibfnamefont {N.}~\bibnamefont
  {Alidoust}}, \bibinfo {author} {\bibfnamefont {H.}~\bibnamefont {Zheng}},
  \bibinfo {author} {\bibfnamefont {M.}~\bibnamefont {Neupane}},  \emph
  {et~al.},\ }\href@noop {} {\bibfield  {journal} {\bibinfo  {journal}
  {Physical Review B}\ }\textbf {\bibinfo {volume} {92}},\ \bibinfo {pages}
  {235104} (\bibinfo {year} {2015})}\BibitemShut {NoStop}%
\bibitem [{\citenamefont {Sun}\ \emph {et~al.}(2016)\citenamefont {Sun},
  \citenamefont {Zhang}, \citenamefont {Felser},\ and\ \citenamefont
  {Yan}}]{sun2016strong}%
  \BibitemOpen
  \bibfield  {author} {\bibinfo {author} {\bibfnamefont {Y.}~\bibnamefont
  {Sun}}, \bibinfo {author} {\bibfnamefont {Y.}~\bibnamefont {Zhang}}, \bibinfo
  {author} {\bibfnamefont {C.}~\bibnamefont {Felser}}, \ and\ \bibinfo {author}
  {\bibfnamefont {B.}~\bibnamefont {Yan}},\ }\href@noop {} {\bibfield
  {journal} {\bibinfo  {journal} {Physical Review Letters}\ }\textbf {\bibinfo
  {volume} {117}},\ \bibinfo {pages} {146403} (\bibinfo {year}
  {2016})}\BibitemShut {NoStop}%
\bibitem [{\citenamefont {Fang}\ \emph {et~al.}(2016)\citenamefont {Fang},
  \citenamefont {Weng}, \citenamefont {Dai},\ and\ \citenamefont
  {Fang}}]{fang2016topological}%
  \BibitemOpen
  \bibfield  {author} {\bibinfo {author} {\bibfnamefont {C.}~\bibnamefont
  {Fang}}, \bibinfo {author} {\bibfnamefont {H.}~\bibnamefont {Weng}}, \bibinfo
  {author} {\bibfnamefont {X.}~\bibnamefont {Dai}}, \ and\ \bibinfo {author}
  {\bibfnamefont {Z.}~\bibnamefont {Fang}},\ }\href@noop {} {\bibfield
  {journal} {\bibinfo  {journal} {Chinese Physics B}\ }\textbf {\bibinfo
  {volume} {25}},\ \bibinfo {pages} {117106} (\bibinfo {year}
  {2016})}\BibitemShut {NoStop}%
\bibitem [{\citenamefont {Blaha}\ \emph {et~al.}(2001)\citenamefont {Blaha},
  \citenamefont {Schwarz}, \citenamefont {Madsen}, \citenamefont {Kvasnicka},
  \citenamefont {Luitz} \emph {et~al.}}]{2001wien2k}%
  \BibitemOpen
  \bibfield  {author} {\bibinfo {author} {\bibfnamefont {P.}~\bibnamefont
  {Blaha}}, \bibinfo {author} {\bibfnamefont {K.}~\bibnamefont {Schwarz}},
  \bibinfo {author} {\bibfnamefont {G.~K.}\ \bibnamefont {Madsen}}, \bibinfo
  {author} {\bibfnamefont {D.}~\bibnamefont {Kvasnicka}}, \bibinfo {author}
  {\bibfnamefont {J.}~\bibnamefont {Luitz}},  \emph {et~al.},\ }\href@noop {}
  {\bibfield  {journal} {\bibinfo  {journal} {An augmented plane wave+ local
  orbitals program for calculating crystal properties}\ }\textbf {\bibinfo
  {volume} {60}} (\bibinfo {year} {2001})}\BibitemShut {NoStop}%
\end{thebibliography}%
\bibliographystyle{apsrev4-1}

\end{document}